\documentclass[12pt]{article}

\usepackage{amsfonts,amssymb}
\newcommand{\bbaa}{\begin{eqnarray}}
\newcommand{\eeaa}{\end{eqnarray}}
\newcommand{\mbb }[1]{\mathbb #1}
\newcommand{\pp}[1]{(\ref{#1})}
\newcommand{\nn}{\nonumber}

\begin{document}
\newpage
\setcounter{page}{0} \thispagestyle{empty}
\begin{flushright}
{nlin.SI/0506041}\\
\end{flushright}
\vfill

\begin{center}
{\LARGE {\bf  Fermionic one- and two-dimensional
}}\\[0.3cm]
{\LARGE {\bf Toda  lattice  hierarchies }}
\\[0.3cm]
{\LARGE {\bf
 and their bi-Hamiltonian structures}}\\[1.4cm]

{\large V.V. Gribanov$^{1}$, V.G. Kadyshevsky,$^{2}$ and A.S.
Sorin$^{3}$}
{}~\\
\quad \\

{{\em Bogoliubov Laboratory of Theoretical Physics,}}\\
{{\em Joint Institute for Nuclear Research,}}\\
{\em 141980 Dubna, Moscow Region, Russia}~\quad\\
{}~
\end{center}

\vfill

\centerline{{\bf Abstract}} \noindent
   By exhibiting the corresponding Lax pair representations we propose a wide class
of integrable two-dimensional (2D) fermionic Toda lattice (TL)
hierarchies which includes the 2D $N=(2|2)$ and $N=(0|2)$
supersymmetric TL hierarchies as particular cases. Performing
their reduction to the one-dimensional case by imposing suitable
constraints we derive the corresponding 1D fermionic TL
hierarchies. We develop the generalized graded R-matrix formalism
using the generalized graded bracket on the space of graded
operators with an involution generalizing the graded commutator in
superalgebras, which allows one to describe these hierarchies in
the framework of the Hamiltonian formalism and construct their
first two Hamiltonian structures. The first Hamiltonian structure
is obtained for both bosonic and fermionic Lax operators while
the second Hamiltonian structure is established for bosonic Lax
operators only. We propose the graded modified Yang-Baxter
equation in the operator form and demonstrate that for the class
of graded antisymmetric $R$-matrices it is equivalent to the
tensor form of the graded classical Yang-Baxter equation.

{}~



\vfill

\vspace{1.5cm}

\begin{flushleft}
{\em ~~~E-mail:}\\
{\em 1) gribanov@thsun1.jinr.ru}\\
{\em 2) kadyshev@jinr.ru}\\
{\em 3) sorin@thsun1.jinr.ru}
\end{flushleft}

\thispagestyle{empty}
 ~\

~\

\newpage
\pagenumbering{arabic}

\section{Introduction}
 The Toda lattice (TL) is one of the most
important families of models in the theory of integrable systems.
Its various generalizations and supersymmetric extensions, having
deep implications in modern mathematical physics, have been the
subject of intense investigations during the last decades.

 The 2D TL hierarchy was first studied in
\cite{Mikh,UT}, and at present two different nontrivial
supersymmetric extensions of 2D TL are known. They are the
$N=(2|2)$ \cite{Olsh}-\cite{DLS} and $N=(0|2)$ \cite{DLS,KS1}
supersymmetric TL hierarchies that possess a different number of
supersymmetries and contain the $N=(2|2)$ and $N=(0|2)$ TL
equations as subsystems. Quite recently, the 2D generalized
fermionic TL equations have been introduced \cite{DDNS} and their
two reductions related to the $N=(2|2)$ and $N=(0|2)$
supersymmetric TL equations were considered. In the present paper,
we describe  a wide class of integrable two-dimensional fermionic
Toda lattice  hierarchies -- 2D fermionic $(K^+,K^-)$-TL
hierarchies, which includes the 2D $N=(2|2)$ and $N=(0|2)$
supersymmetric TL hierarchies  as particular cases and contains
 the 2D generalized
fermionic TL equations as a subsystem.

The Hamiltonian description of the 2D TL hierarchy has been
constructed only quite recently
 in the framework of the R-matrix approach  in \cite{Carlet}, where the
 new R-matrix associated with splitting of
 algebra given by a pair of difference operators was introduced.
 In the
present paper, we adapt this R-matrix to the case of $Z_2$-graded
operators and derive the bi-Hamiltonian structure of the 2D
fermionic $(K^+,K^-)$-TL hierarchy.

Remarkably,  in solving this problem the generalized graded
bracket \pp{SK-bracket} on the space of graded  operators with an
involution  finds its new application. This bracket was
 introduced in \cite{KS2}, where it was observed that the
$N=(1|1)$ supersymmetric 2D TL hierarchy had a  natural Lax-pair
representation in terms of this bracket which allowed one to
derive  the dispersionless $N=(1|1)$ 2D TL hierarchy and its  Lax
representation.  In the present paper, the generalized graded
bracket is used to describe the 2D fermionic $(K^+,K^-)$-TL
hierarchy and define its two Hamiltonian structures.
 Moreover, we
demonstrate that the classical graded Yang-Baxter equation
\cite{KulSk} has an equivalent operator representation in terms of
the generalized graded bracket. All these facts attest  that this
bracket has a fundamental meaning  and  allows a broad spectrum of
applications  in modern mathematical physics.

This paper is the extended version of \cite{GKS}. The structure of
the paper is as follows.

In Sec. 2, we define the space of the $Z_2$-graded difference
operators with the involution and recall the generalized graded
bracket \cite{KS2} and its  properties.

In Sec.~3, we give a theoretical background of the $R$-matrix
method generalized to the case  of the $Z_2$-graded difference
operators. We define the $R$-matrix on the associative algebra
{\sf g} of the $Z_2$-graded difference operators, derive the
graded modified  Yang-Baxter equation and using the generalized
graded bracket obtain two Poisson brackets for the functionals on
${\sf g}^\dag={\sf g}$. The proper properties of the Poisson
brackets thus obtained are provided by the properties of the
generalized graded bracket. Thus, for the $Z_2$-graded difference
operators of odd (even) parity this bracket defines odd (even)
first Poisson bracket. The second Poisson bracket is found only
for even difference operators which in this case are compatible
with the first Poisson bracket. Using these Poisson brackets  one
can define the Hamiltonian equations that can equivalently be
rewritten  in terms of the Lax-pair representation. The basic
results of Sec. 3 are formulated as Theorem.

In Sec.~4, using the generalized graded bracket we propose a new
2D fermionic $(K^+,K^-)$-TL hierarchy in terms of the Lax-pair
representation and construct the algebra of its  flows. Then we
present the explicit expression for its flows and show that all
known up to now 2D TL equations can be derived from this hierarchy
as subsystems.

 In Sec. 5, we consider the reduction of the 2D
fermionic $(K^+,K^-)$-TL hierarchy to the 1D space and reproduce
the 1D generalized fermionic TL equations \cite{DDNS} as the first
flow of the reduced hierarchy with additional constraint imposed.

In Secs. 6 and 7, we apply the results of Sec. 3 to derive the
Hamiltonian structures of the 1D and 2D fermionic $(K^+,K^-)$-TL
hierarchies. Following \cite{Carlet} we use the $R$-matrix which
acts nontrivially on the space of the direct sum of two difference
operators and derive two different Hamiltonian structures of the
2D fermionic $(K^+,K^-)$-TL hierarchy. The first Hamiltonian
structure is obtained for both even and odd values of $(K^+,K^-)$
while the second one is found  for even values of $(K^+,K^-)$
only. We perform their  Dirac reduction  and demonstrate that in
general the Dirac brackets for the second Hamiltonian structure
are nonlocal but for the  case of the fermionic $(2,2)$-TL
hierarchy they become local. As an example, we give the explicit
form of the first and second Hamiltonian structures for the
fermionic 1D (2,2)-TL hierarchy.

In Sec.~8, we briefly summarize the main results obtained in this
paper and point out open problems. In Appendices, we clarify some
technical aspects.

\section{ Space of difference operators}
 In this section, we define the space of difference operators
which will play an important role in our consideration. These
operators can be represented in the following general form:
 \bbaa  \label{O}
\mbb{O}_m=\sum\limits_{k=-\infty}^{\infty}f^{(m)}_{k,j}\ e^{(k-m)
\partial}, \ \ \  \ \ \ \ m, j \in \mathbb{Z},
 \eeaa
 parameterized
by the functions $f_{2k,j}^{(m)}$  ($f_{2k+1,j}^{(m)}$) which are
the $Z_2$-graded bosonic (fermionic) lattice fields with the
lattice index $j$ $(j\in\mbb{Z})$ and the Grassmann parity defined
by index~$k$
\bbaa \label{f-grad} d_{f^{(m)}_{k,j}}=|k|\ \mbox{mod}\ 2.\nn
\eeaa
%
%
%
 $e^{k
\partial}$ is the shift operator whose action on the lattice fields
results into a discrete shift of a lattice index
 \bbaa \label{shift}
 e^{l \partial}f^{(m)}_{k,j}=f^{(m)}_{k,j+l}e^{l \partial}.
 \eeaa
The shift operator has  $Z_2$-parity defined as
\bbaa d_{e^{l\partial}}'=|l|\ \mbox{mod} \ 2.\nn \eeaa
The operators  $ \mbb{O}_m$ \pp{O} admit the diagonal $Z_2$-parity
\bbaa\label{Z2-par}
d_{\mbb{O}_m}=d_{f_{k,j}^{(m)}}+d_{e^{(k-m)\partial}}'=|m|\
\mbox{mod} \ 2 \eeaa
 and the involution
\bbaa
\mbb{O}^*_m=\sum\limits_{k=-\infty}^{\infty}(-1)^{k}f_{k,j}^{(m)}\
e^{(k-m)
\partial}.\nn\eeaa

 In what follows we also need the projections
 of the operators $\mbb{O}_m$ defined as
\bbaa  ( \mbb{O}_m)_{\leqslant p}=\sum\limits_{k\leqslant
p+m}f_{k,j}\ e^{(k-m)
\partial},\ \ \ \
 (\mbb{O}_m)_{\geqslant p}=\sum\limits_{k\geqslant p+m}f_{k,j}\
 e^{(k-m)
\partial}\nn
 \eeaa
and we will use the usual notation for the projections
$(\mbb{O}_m)_+:=(\mbb{O}_m)_{\geqslant 0}$ and
$(\mbb{O}_m)_-:=(\mbb{O}_m)_{< 0}$.
 Note that $e^{l\partial}$  is a conventional form for the
shift operators defined in terms of infinite-dimensional matrices
 $(e^{l\partial})_{i,j}\equiv\delta_{i,j-l}$, and
there is an isomorphism between operators \pp{O} and
infinite-dimensional matrices (see e.g. \cite{AB})
\bbaa \mbb{O}_m=\sum\limits_{k=-\infty}^{\infty}f_{k,j}^{(m)}\
e^{(k-m)
\partial}\ \ \to \ \ {(\mbb{O}_m)}_{j,i}\equiv\sum\limits_{k=-\infty}^{\infty} f_{k,j}^{(m)}\
\delta_{j,i-k+m}.\nn \eeaa

In the operator space \pp{O} one can extract two subspaces which
are of great importance  in our further consideration
\bbaa  \label{O+-}
\mbb{O}^\pm_{K^\pm}&=&\sum\limits_{k=0}^{\infty}f_{k,j}^\pm\
^{\pm(K^\pm-k)
\partial},\ \ \ \  \ K^\pm\in\mathbb{N}.
%
%
 \eeaa
 \looseness=-1
The operators of  the subspaces $ \mbb{O}^\pm_{K^\pm}$
 form associative algebras with the
multiplication \pp{shift}. Using this fact we define on these
subspaces the generalized graded  algebra with the bracket
\cite{KS2}
\bbaa\label{SK-bracket}  [{\mathbb O}, \widetilde{ \mathbb O}
\}:={\mathbb O}\ \widetilde{\mathbb O}
 - (-1)^
 {d_{{\mathbb O}
 \vphantom{\widetilde{\mathbb O}} }
 d_{\widetilde{\mathbb O}}}~{\widetilde{\mathbb O}}^{*(d_{{\mathbb
O}\vphantom{\widetilde{\mathbb O}}})}~ {{\mathbb
O}}^{*(d_{\widetilde{\mathbb O}})}, \eeaa
 where the operators ${{\mathbb O}}$  and  ${\widetilde{\mathbb O}}$ belong to the
 subspaces $\mbb{O}^+_{K^+}\ (\mbb{O}^-_{K^-})$, and
${{\mathbb O}}^{*(m)}$  denotes the $m$-fold action of the
involution $*$ on the operator ${\mathbb O}$,  (${{\mathbb
O}}^{*(2)}={\mathbb O}$). Bracket \pp{SK-bracket} generalizes  the
(anti)commutator in superalgebras and satisfies the following
properties \cite{KS2}:

symmetry
 \bbaa \label{symSK} [ {\mathbb O}, \widetilde{ \mathbb O}
\}=
 - (-1)^
 {d_{{\mathbb O}
 \vphantom{\widetilde{\mathbb O}} }
 d_{\widetilde{\mathbb O}}}~[{\widetilde{\mathbb O}}^{*(d_{{\mathbb
O}\vphantom{\widetilde{\mathbb O}}})},  {{\mathbb
O}}^{*(d_{\widetilde{\mathbb O}})}\}, \eeaa

 derivation
 \bbaa \label{derSK}
 [{\mathbb O}, \widetilde{ \mathbb O}\, \widehat{ \mathbb O}
\}=[ {\mathbb O}, \widetilde{ \mathbb O}\}\, \widehat{ \mathbb O}
 + (-1)^{d_{{\mathbb O}
 \vphantom{\widetilde{\mathbb O}} } d_{\widetilde{\mathbb O}}}\,
 {\widetilde{\mathbb
O}}^{*(d_{{\mathbb O}\vphantom{\widetilde{\mathbb O}}})} [{\mathbb
O}^{*(d_{\widetilde{\mathbb O}})}, {\widehat{\mathbb O}}\},\eeaa

 and  Jacobi identity
\bbaa\label{SK-Jacobi} (-1)^{d_{{\mathbb O}
 \vphantom{\widetilde{\mathbb O}} } d_{\widehat{\mathbb O}\vphantom{\widetilde{\mathbb
 O}}}}\,
 [[{{\mathbb O}},\, {\widetilde{\mathbb O}}^{*(d_{{\mathbb
O}\vphantom{\widetilde{\mathbb O}}})}\},\,  {\widehat{\mathbb
O}}^{*(d_{{\mathbb O}\vphantom{\widetilde{\mathbb O}}}+
d_{\widetilde{\mathbb O}}) \vphantom{\widetilde{\mathbb O}}}\}
 +(-1)^{d_{\widetilde{{\mathbb O}}
 \vphantom{\widetilde{\mathbb O}} } d_{{\mathbb O}\vphantom{\widetilde{\mathbb
 O}}}}\,
 [[{\widetilde{\mathbb O}},\, {\widehat{\mathbb O}}^{*(d_{\widetilde{{\mathbb
O}}\vphantom{\widetilde{\mathbb O}}})}\},\,
 {{\mathbb O}}^{*(d_{\widetilde{{\mathbb O}}\vphantom{\widetilde{\mathbb
O}}}+d_{\widehat{\mathbb O}}) \vphantom{\widetilde{\mathbb
O}}}\}&&\nn\\
 +\ \  (-1)^{d_{\widehat{\mathbb O}
 \vphantom{\widetilde{\mathbb O}} } d_{\widetilde{\mathbb O}\vphantom{\widetilde{\mathbb
 O}}}}\,
[[{\widehat{\mathbb O}},\, {{\mathbb O}}^{*(d_{\widehat{\mathbb
O}\vphantom{\widetilde{\mathbb O}}})}\},\,
 {\widetilde{\mathbb O}}^{*(d_{\widehat{{\mathbb O}}\vphantom{\widetilde{\mathbb
O}}}+d_{{\mathbb O}}) \vphantom{\widetilde{\mathbb O}}}\}
 &=&0.
\eeaa

For the operators $\mathbb{O}_m$ \pp{O} we define the supertrace
\bbaa \label{supertr}
 str \mathbb O=\sum\limits_{j=-\infty}^{\infty}(-1)^j
f^{(m)}_{m,j}. \eeaa
In what follows we assume  suitable boundary conditions for the
functions $f_{k,j}^{(m)}$ in order the main property of
supertraces
\begin{eqnarray}\label{suptr0}
str [ {\mathbb O}, \widetilde{\mathbb O} \}=0
\end{eqnarray}
 be satisfied for the case of the
generalized graded bracket  \pp{SK-bracket}.

\section{R-matrix formalism}
 In this section, we develop  a
theoretical background of the R-matrix method adapted to the case
of the operator space \pp{O}.

 Let {\sf g} be an associative algebra of the operators from the space \pp{O} with
the invariant non-degenerate inner product
\bbaa \label{InProd}
<\mbb{O},\widetilde{\mbb{O}}>=str(\mbb{O}\,\widetilde{\mbb{O}})\nn
\eeaa
using which one can identify the algebra {\sf g} with its dual
{\sf g${}^\dag$}.  We set the following Poisson bracket:
\bbaa \label{pb1} \{f,g\}(\mbb{O})=-<\mbb{O},[\nabla g,(\nabla
f)^{*(\nabla d_{g})}\} >, \eeaa
where $f, g$ are functionals on {\sf g},
and $\nabla f$ and  $\nabla g$ are their gradients at the point
$\mbb{O}$
 which are
related with $f,g $ through  the inner product
\bbaa \frac{\partial f(\mbb{O}+\epsilon \delta
\mbb{O})}{\partial\epsilon}{\Biggr |}_{\epsilon=0} =\ <\delta
\mbb{O}, \nabla f(\mbb{O})>. \nn\eeaa
%
%
%
Note that the proper properties of the Poisson bracket \pp{pb1}
follow from the properties (\ref{symSK}--\ref{SK-Jacobi}) of the
generalized bracket \pp{SK-bracket} and are strictly determined by
the $Z_2$-parity of the operator $\mathbb{O}$. Thus, one has
 symmetry
 \bbaa\label{sym}
\{f,g\}&=&-(-1)^{(d_f+d_\mathbb{O})(d_g+d_\mathbb{O})}\{g,f\},
\eeaa
 derivation
\bbaa \label{der} \{f,g h\}&=&
\{f,g\}h+(-1)^{d_g (d_f+d_\mathbb{O})} g\{f,h\},
 \eeaa
%
%
and Jacobi identity
 \bbaa\label{Jac}
(-1)^{(d_f+d_\mathbb{O})(d_h+d_\mathbb{O})}\{\{f,g\},h\}
+(-1)^{(d_g+d_\mathbb{O})(d_f+d_\mathbb{O})}\{\{g,h\},f\}\nn\\
 +(-1)^{(d_h+d_\mathbb{O})(d_g+d_\mathbb{O})}\{\{h,f\},g\}&=&0. \eeaa
Therefore, for the even operator  $\mathbb{O}$ one has a usual
(even) $Z_2$-graded Poisson bracket, while for the operators with
odd diagonal parity $d_\mathbb{O}$ \pp{Z2-par}  eq. \pp{pb1}
defines the odd $Z_2$-graded Poisson bracket (antibracket).

Having defined the   Poisson bracket   we proceed with the search
for the hierarchy of  flows generated by this  bracket using
Hamiltonians. Therefore, we need to determine an infinite set of
functionals which should be in involution to play the role of
Hamiltonians. For Poisson bracket \pp{pb1} one can find
 an infinite set of Hamiltonians in a rather standard way
\bbaa \label{hamB} H_k=\frac1k
str\mbb{O}^k_*=\frac1k\sum_{i=-\infty}^\infty(-1)^i
f_{km,i}^{(km)}, \eeaa
where $ \mbb{O}^k_*$ is defined as
\bbaa \label{compLax}(\mbb{O})^{2k}_*:=
(\mbb{O}^{*(d_\mbb{O})}\mbb{O})^k, \ \ \ (\mbb{O})^{2k+1}_*:=
\mbb{O}\ (\mbb{O})^{2k}_* .\eeaa
 For the odd operators $\mbb{O}$ eq. \pp{hamB} defines only
fermionic nonzero functionals $H_{2k+1}$, since in this case  even
powers of the operators $\mbb{O}$ have the following
representation:
\bbaa \label{evenH} d_{\mathbb{O}}=1:\ \ \ \ (\mathbb{O})^{2k}_*=
(1/2[(\mathbb{O})^{*},\mathbb{O}\})^k \equiv
1/2[({(\mathbb{O})^{2k-1}_*})^{*},\mathbb{O}\}  \eeaa
and all the bosonic Hamiltonians are trivial ($H_{2k}=0$ ) like
the supertrace of the generalized graded bracket.

  The  functionals \pp{hamB}
 are obviously in involution but produce a trivial dynamics.
 Actually, the functionals $H_k$
\pp{hamB}  are the Casimir operators of the Poisson bracket
\pp{pb1}, so the Poisson bracket of $H_k$ with any other
functional is equal to zero as an output (due to the relation
$\nabla H_{k+1}=\mbb{O}_*^k$). Nevertheless, it is possible to
 modify  the Poisson bracket \pp{pb1} in such a way that the new Poisson
bracket would produce nontrivial  equations of motion using the
same Hamiltonians \pp{hamB}  and these Hamiltonians  are  in
involution with respect to the modified Poisson bracket as well.
Let us introduce the modified generalized  graded bracket on the
space \pp{O}
\bbaa\label{SKmod}
[\mbb{O},\widetilde{\mbb{O}}\}_R:=[R(\mbb{O}),\widetilde{\mbb{O}}\}+[\mbb{O},R(\widetilde{\mbb{O}})\},
\eeaa
where the $R$-matrix
is a linear map $R$: {\sf g $\to$ g}  such that the  bracket
\pp{SKmod} satisfies  the properties
(\ref{symSK}--\ref{SK-Jacobi}). One can verify that the Jacobi
identities \pp{SK-Jacobi} for the bracket \pp{SKmod} can
equivalently be
 rewritten in terms of the generalized graded bracket
\pp{SK-bracket}
\bbaa  (-1)^{d_{{\mathbb O}
 \vphantom{\widetilde{\mathbb O}} } d_{\widehat{\mathbb O}\vphantom{\widetilde{\mathbb
 O}}}}\,
 [[{{\mathbb O}},\, {\widetilde{\mathbb O}}^{*(d_{{\mathbb
O}\vphantom{\widetilde{\mathbb O}}})}\}_R,\, {\widehat{\mathbb
O}}^{*(d_{{\mathbb O}\vphantom{\widetilde{\mathbb O}}}+
d_{\widetilde{\mathbb O}}) \vphantom{\widetilde{\mathbb
O}}}\}_R+\mbox{cycle
perm.}=~~~~~~~~~~~~~~~~~~~~~~~~~~~~~~\nn\\
(-1)^{d_{{\mathbb O}
 \vphantom{\widetilde{\mathbb O}} } d_{\widehat{\mathbb O}\vphantom{\widetilde{\mathbb
 O}}}}\, [R( [{{\mathbb O}},\, {\widetilde{\mathbb O}}^{*(d_{{\mathbb
O}\vphantom{\widetilde{\mathbb O}}})}\}_R )- [R({{\mathbb O}}),\,
R({\widetilde{\mathbb O}}^{*(d_{{\mathbb
O}\vphantom{\widetilde{\mathbb O}}})})\},{\widehat{\mathbb
O}}^{*(d_{{\mathbb O}\vphantom{\widetilde{\mathbb O}}}+
d_{\widetilde{\mathbb O}}) \vphantom{\widetilde{\mathbb
O}}}\}+\mbox{cycle perm.}=0.\nn
 \eeaa
Thus, one can conclude that a sufficient condition for $R$ to be
the $R$-matrix is the validity of the following equation:
\bbaa \label{YB} R([\mbb{O},\widetilde{\mbb{O}}\}_R)-
[R(\mbb{O}),R(\widetilde{\mbb{O}})\}=\alpha
[\mbb{O},\widetilde{\mbb{O}}\}, \eeaa
where $\alpha$ is an arbitrary constant. One can show (see
Appendix A) that for the case of graded antisymmetric operators
$R$
 eq. \pp{YB} at $\alpha=1$ represents the operator form of the
graded classical Yang-Baxter equation \cite{KulSk}.
 Following the terminology
of \cite{STSh} we call
 eq. \pp{YB} the graded  modified  Yang-Baxter equation.
 Equation \pp{YB} is the
generalization of the    graded modified classical Yang-Baxter
equation discussed in paper \cite{Yung} for the case of space of
graded operators \pp{O}.

With the new bracket \pp{SKmod}  one can define the corresponding
new Poisson bracket on dual~${\sf g}^\dag$:
\bbaa \label{PB1} \{f,g\}_1(\mbb{O})&=&-\ 1/2<\mbb{O},[\nabla
g,(\nabla f)^{*( d_{\nabla g})}\}_R >
\nn\\
&\equiv &\frac{1}{2}<(-1)^{d_{\nabla g}d_\mathbb{O}}R(\nabla
g)^{*(d_\mathbb{O})}[\mbb{O}^{*( d_{ \nabla g})},(\nabla f)^{*(
d_{\nabla g})}\}\nn\\
&& -\ [\mbb{O},\nabla g\}R((\nabla f)^{*(d_{\nabla g})})>.~~~~~
 \eeaa
With respect to the dependence of the r.h.s of \pp{PB1} on the
point $\mathbb{O}$ this is a linear bracket. Without going into
details we introduce also bi-linear bracket for bosonic graded
operators $\mathbb{O}_B$  ( $d_{\mathbb{O}_B}=0$) as follows:
\bbaa\label{PB2} \{f,g\}_2(\mathbb{O}_B)&=&-\
1/4<[\mathbb{O}_B,\nabla
g\}R((\nabla f)^{*(d_{\nabla g})} \mathbb{O}_B^{*(d_{\nabla f}+d_{\nabla g})}\nn\\
&&+\ \mathbb{O}_B^{*(d_{\nabla  g})}(\nabla f)^{*(d_{\nabla g})})
-R(\nabla g \mathbb{O}_B^{*(d_{\nabla g})}\nn\\
&&+\ \mathbb{O}_B\nabla g)[\mathbb{O}_B^{*(d_{\nabla  g})},(\nabla
f)^{*( d_{\nabla g})}\}>.~~~~  \eeaa
We did not succeed in constructing the bi-linear bracket for the
case of fermionic operators $\mathbb{O}_F$ (
$d_{\mathbb{O}_F}=1$). The bracket \pp{PB1} is obviously the
Poisson bracket if $R$ is an $R$-matrix on ${\sf g}$. The
bi-linear bracket \pp{PB2} becomes Poisson bracket under more
rigorous constraints which can be found in the following
%
%

{\bf ~~Theorem.}\ a) Linear bracket \pp{PB1} is the Poisson
bracket if $R$ obeys the graded modified Yang-Baxter equation \pp{YB};\\
b) the bi-linear bracket \pp{PB2} is the Poisson bracket if $R$
and its graded antisymmetric part $1/2(R-R^\dag)$ obey the graded
modified
Yang-Baxter equation \pp{YB} with the same $\alpha$;\\
%
%
c) if $\mathbb{O}=\mathbb{O}_B$, then these two Poisson brackets
are compatible;\\
%
%
%
%
d) the  Casimir operators $H_{k}$ \pp{hamB}  of the bracket
\pp{pb1} are in involution with respect to both linear \pp{PB1}
and
bi-linear \pp{PB2} Poisson brackets;\\
e) the Hamiltonians $H_{k}\neq 0$  \pp{hamB} generate evolution
equations
\bbaa\label{Lax-R}
\partial_k \mathbb{O}&=&\{H_{k+1},\mathbb{O}\}_1=1/2 [R((\nabla H_{k+1})^{*(d_\mathbb{O})}),\mathbb{O}\},\nn\\
\partial_k \mathbb{O}_B&=&\{H_k,\mathbb{O}_B\}_2=1/4 [R(\nabla H_k \mathbb{O}_B+\mathbb{O}_B\nabla
H_k),\mathbb{O}_B\}\eeaa
via the brackets \pp{PB1} and \pp{PB2}, respectively, which
connect the Lax-pair and Hamiltonian representations. ~~

{\bf Proof.}\ a). This is a summary of the above discussion on the
linear
bracket.\\
b). Using the property of cyclic permutations inside the
supertrace one can easily verify that the symmetry property
$\{f,g\}_2=-(-1)^{d_f d_g} \{g,f\}_2$ holds. Verification of  the
Jacobi identities for the bi-linear bracket amounts to
straightforward and tedious calculations which are presented in
Appendix B.\\
c). These two Poisson brackets are obviously compatible. Indeed, a
deformation of the point $\mathbb{O}_B\longrightarrow
\mathbb{O}_B+b$ on the dual $\sf g^\dag$, where $b$ is an
arbitrary constant operator, transforms \pp{PB2} into the sum of
two Poisson brackets
\bbaa \{f,g\}_2(\mathbb{O}_B+b)=\{f,g\}_2(\mathbb{O}_B)+b
\{f,g\}_1(\mathbb{O}_B).\nn \eeaa
d). Substituting the expressions for  Hamiltonians \pp{hamB} into
eqs. \pp{PB1} and \pp{PB2} and taking into account that $\nabla
H_{k+1}=\mathbb{O}^k_*$ it is easily to check that the Casimirs of
the bracket \pp{pb1} are in involution with respect
to both the Poisson structures \pp{PB1} and \pp{PB2}.\\
 e).  Using cyclic permutations inside the supertrace
operation let us rewrite both the Poisson brackets \pp{PB1} and
\pp{PB2} in the following general form:
 \bbaa \{f,g\}_i(\mathbb{O})=<P_i(\mathbb{O}) \nabla g,(\nabla f)^{*(d_{\nabla g})}>,\ \
 \ \ i=1,2,\nn
 \eeaa
 where $P_i(\mathbb{O})$ is the Poisson tensor corresponding to the bracket
$ \{..,..\}_i$
\bbaa P_1(\mathbb{O})\nabla g&=&-1/2 ([\mathbb{O},R(\nabla g)\}
+R^\dag([\mathbb{O},\nabla g]\})),\nn \\
 P_2(\mathbb{O}_B) \nabla g&=&1/4 ([R(\nabla g
\mathbb{O}_B+\mathbb{O}_B \nabla g),\mathbb{O}_B^{*(d_{\nabla g})}\}\nn\\
&-&\mathbb{O}_BR^\dag([\mathbb{O}_B,\nabla
g\})-R^\dag([\mathbb{O}_B,\nabla g\})\mathbb{O}_B^{*(d_{\nabla
g})} ) \nn \eeaa
and the adjoint operator $R^\dag$ acts on the dual $\sf g^\dag$
$$<\mathbb{O},R(\widetilde{\mathbb{O}})>=<R^\dag(\mathbb{O}),\widetilde{\mathbb{O}}>.$$
The Hamiltonian vector field associated with Hamiltonian $H_k$ is
given by $\partial_k \mathbb{O}=P_i(\mathbb{O})\nabla H_k$.
%
%
%
%
%
%
Taking into account that  $[\mathbb{O},\nabla H_k\}=0$ we arrive
at the Lax-pair representations \pp{Lax-R}.
$~~~~~\blacksquare$

 Note that a similar Theorem  when  the shift
operators and  functions parameterizing the difference operators
$\mathbb{O}$ \pp{O} have  even $Z_2$-parity  was discussed in
\cite{STSh,OR,LCP,Carlet}.

For the  graded modified Yang-Baxter equation \pp{YB} there is a
particular class of solutions which are useful in application.
Suppose that the algebra {\sf g} can be represented as a vector
space direct sum of two subalgebras
$${\sf g=g{}_+\dotplus g{}_-}:\ \ \ \ \
 [{\sf g}_+,{\sf g}_+\}\subset {\sf g}_+, \ \ \ \ [{\sf
g}_-,{\sf g}_-\}\subset {\sf g}_-.$$
 Let $P_\pm$ be the projection
operators on these subalgebras, $P_\pm{\sf g}={\sf g}_\pm$, then
one can easily verify that $R=P_+-P_-$ satisfies the graded
modified
 Yang-Baxter equation \pp{YB} at $\alpha=1$ and, therefore,
represents the $R$-matrix on {\sf g}. Indeed, in this case the
 modified generalized graded bracket \pp{SKmod}
\bbaa [\mbb{O},\widetilde{\mbb{O}}\}_R=2
[(\mbb{O})_+,(\widetilde{\mbb{O}})_+\}-2[(\mbb{O})_-,(\widetilde{\mbb{O}})_-\}
\eeaa
 obviously satisfies the Jacobi identities \pp{SK-Jacobi}, since
it determines the usual direct sum of two subalgebras $$[{\sf
g}_\pm,{\sf g}_\pm\}_R\subset {\sf g}_\pm,\ \ \ \ \ \ [{\sf
g}_+,{\sf g}_-\}_R=0.$$

\section{ 2D  fermionic $(K^+,K^-)$-Toda lattice hierarchy}
 In this section, we introduce the
two-dimensional fermionic $(K^+,K^-)$-Toda lattice hierarchy in
terms of the Lax-pair representation.

 Let us consider two difference operators $L^\pm_{K^\pm}$
 \bbaa\label{LaxOp} L^+_{K^+}=\sum\limits_{k=0}^\infty u_{k,i}
e^{(K^+-k)\partial}, \ \ \ \ \ \ \
L^-_{K^-}=\sum\limits_{k=0}^\infty v_{k,i} e^{(k-K^-)\partial},
 \eeaa
 which obviously belong to the spaces \pp{O+-}.
 The lattice fields   and the shift operator entering
into these operators have the following length dimensions:
$[u_{k,i}]=-1/2 k$,
 $[v_{k,i}]=1/2(k-K^+-K^-)$  and $[e^{k\partial}]=-1/2k$,
 respectively, so  operators
 \pp{LaxOp} are of equal length dimension,
 $[L^+_{K^+}]=[L^-_{K^-}]=-1/2 K^+$.
The dynamics of the fields $u_{k,i},v_{k,i}$ are governed by the
Lax equations expressed in terms of the generalized graded bracket
\pp{SK-bracket} \cite{KS2}
\begin{eqnarray}\label{LaxEq}
\ D_s^\pm L^\alpha_{K^\alpha}=\mp \alpha (-1)^{s K^\alpha K^\pm
}[(((L^\pm_{K^\pm})_*^s)_{-\alpha})^{*(K^\alpha)},
L^\alpha_{K^\alpha}\},\ \ \ \ \
%
 ~ \alpha=+,-,
\ \ \ s \in \mathbb{N},
 \eeaa
where $D^\pm_s$  are evolution derivatives with the $Z_2$-parity
defined as
$$ d_{D^\pm_s}=sK^\pm\ \mbox{ mod } \ 2
 $$ and the length dimension $[D^+_s]=[D^-_s]= - sK^+/2.$
The Lax equations \pp{LaxEq} generate  non-Abelian  (super)algebra
of flows of the 2D fermionic $(K^+,K^-)$-TL hierarchy
\bbaa [D^\pm_s,D^\pm_p\}=(1-(-1)^{spK^\pm})D^\pm_{s+p},\ \ \ \
 \bigl[ D^+_s,D^-_p\} =   0.\nn
 \eeaa
The composite  operators $(L^\pm_{K^\pm})_*^s$
entering into the
 Lax equations \pp{LaxEq} are defined by eq. \pp{compLax}
  and  also belong to the spaces
 \pp{O+-}
\bbaa\label{LaxPower} (L^+_{K^+})^r_*:=\sum\limits_{k=0}^{\infty}
u_{k,i}^{(r)} e^{(rK^+-k)\partial}, \ \ \ \ \ \ \
(L^-_{K^-})^r_*:=\sum\limits_{k=0}^\infty v_{k,i}^{(r)}
e^{(k-rK^-)\partial}.~~~~~~~\nn \eeaa
Here $u_{k,i}^{(r)}$ and $v_{k,i}^{(r)}$ are functionals of the
original fields
 and there are the  following recursion relations for them
\bbaa\label{CompFields} u_{p,i}^{(r+1)}&=&\sum\limits_{k=0}^p
(-1)^{k K^+} u_{k,i}^{(r)}u_{p-k,i-k+rK^+},\ \ \ \ u_{p,i}^{(1)}=u_{p,i},\nn\\
v_{p,i}^{(r+1)}&=&\sum\limits_{k=0}^{p}(-1)^{k K^-}
v_{k,i}^{(r)}v_{p-k,i+k-rK^-}, \ \ \ \ v_{p,i}^{(1)}=v_{p,i}.
\nn\eeaa
Now using the Lax representation \pp{LaxEq} and relations
\pp{derSK} and \pp{compLax} one can derive the equations of motion
for the composite Lax operators
\begin{eqnarray}\label{LaxEqCom}
\ D_s^\pm (L^\alpha_{K^\alpha})^r_*&=&\mp \alpha (-1)^{s  r
K^\alpha K^\pm }[(((L^\pm_{K^\pm})_*^s)_{-\alpha})^{*(rK^\alpha)},
(L^\alpha_{K^\alpha})^r_*\}.
 \eeaa

%
%
%
 The Lax-pair representation  (\ref{LaxEqCom}) generates the
following equations for the functionals
$u_{k,i}^{(r)},v_{k,i}^{(r)}$:
\bbaa \label{Eqs1} D^+_s u_{k,i}^{(r)}&=&\sum\limits_{p=1}^{k}
((-1)^{rpK^++1}u_{p+sK^+,i}^{(s)}u_{k-p,i-p}^{(r)}
\nn\\
&+&(-1)^{(k+p)sK^+}u_{k-p,i}^{(r)}u_{p+sK^+,i+p-k+rK^+}^{(s)}),\\
\label{Eqs2} D^-_s u_{k,i}^{(r)}&=&\sum\limits_{p=0}^{sK^--1}
((-1)^{(sK^-+p)rK^+}v_{p,i}^{(s)}u_{p+k-sK^-,i+p-sK^-}^{(r)}
\nn\\
&-&(-1)^{(k+p+1)sK^-}u_{p+k-sK^-,i}^{(r)}v_{p,i-p-k+sK^-+rK^+}^{(s)}),\\
\label{Eqs3} D^+_s v_{k,i}^{(r)}&=&\sum\limits_{p=0}^{sK^+}
((-1)^{(sK^++p)rK^-}u_{p,i}^{(s)}v_{p+k-sK^+,i-p+sK^+}^{(r)}
\nn\\
&-&(-1)^{(k+p+1)sK^+}v_{p+k-sK^+,i}^{(r)}u_{p,i+p+k-sK^+-rK^-}^{(s)}),\\
\label{Eqs4}D^-_s v_{k,i}^{(r)}&=&\sum\limits_{p=0}^{k}
((-1)^{rpK^-+1}v_{p+sK^-,i}^{(s)}v_{k-p,i+p}^{(r)}\nn\\
\nn\\
&+&(-1)^{(k+p)sK^-}v_{k-p,i}^{(r)}v_{p+sK^-,i+k-p-rK^-}^{(s)}).
\eeaa
It is assumed that in the right-hand side of  eqs.
(\ref{Eqs1}--\ref{Eqs4})
 all the
functionals $u_{k,i}^{(r)}$ $v_{k,i}^{(r)}$ with $k<0$  should be
set equal to zero.

Let us demonstrate that all known up to now 2D supersymmetric Toda
lattice equations can be derived from the system
(\ref{Eqs1}--\ref{Eqs4}).

First, the 2D generalized fermionic Toda lattice equation
discussed   in \cite{DDNS} can be reproduced from the system of
equations (\ref{Eqs1}--\ref{Eqs4}) as a subsystem with additional
reduction constraints imposed. In order to see this, let us
introduce the notation
$v_{0,i}=d_i,v_{1,i}=\rho_i,u_{1,i}=\gamma_i, u_{2,i}=c_i$ and
 consider eqs. (\ref{Eqs2}) and (\ref{Eqs4}) at  $K^+=K^-=2$, $r=s=1$.
One obtains
\bbaa\label{2dToda} &&D_1^+d_i=d_i(c_i-c_{i-2}),\ \ \ \ \ \
D_1^-\gamma_i=\rho_{i}u_{0,i-1}-\rho_{i+2}u_{0,i},
\nn \\
&& D_1^-c_i=d_{i}u_{0,i-2} -d_{i+2}u_{0,i}-\gamma_i
\rho_{i+1}-\gamma_{i-1}\rho_i,\nn\\
&& D_1^+\rho_i=\rho_i(c_i-c_{i-1})+d_{i+1}
\gamma_i-d_i\gamma_{i-2},\ \ \ \ \ \ D_1^-u_{0,i}=0. \eeaa
It is easy to check that after reduction $u_{0,i}=1$ eqs.
(\ref{2dToda}) coincide with the 2D generalized fermionic Toda
lattice equations up to time redefinition $D_1^-\to-D_1^-$.

Next, the N=$(2|2)$ supersymmetric Toda lattice equation also
belongs to the system (\ref{Eqs1}--\ref{Eqs4}). In order to see
that, let us consider eqs. \pp{Eqs2} at $K^+=K^-=s=k=r=1$
\bbaa \label{1}D_1^-u_{1,i}=-u_{0,i-1}v_{0,i}-u_{0,i}v_{0,i+1}
\eeaa
and eqs. \pp{Eqs3} at $K^+=K^-=s=r=1$, $k=0$
\bbaa \label{2}D_1^+v_{0,i}=v_{0,i}(u_{1,i}-u_{1,i-1}). \eeaa
Then imposing the constraint $u_{0,i}=1$ and eliminating the
fields $u_{1,i}$ from eqs. (\ref{1}--\ref{2}) one obtains the
N=$(1|1)$ superfield form of the N=$(2|2)$ supersymmetric Toda
lattice equation \cite{Ik,EH,LSor1,DLS}
\bbaa \label{3} D^+_1D^-_1\mbox{ ln }v_{0,i}=v_{0,i+1}-v_{0,i-1}.
\eeaa
Analogously, one can show that the consideration of eqs. \pp{Eqs3}
and \pp{Eqs4} at $K^+=1$, $K^-=2$, $s=r=1$ and $k=0,1$  leads to
the N=$(0|1)$ superfield form of the N=$(0|2)$ supersymmetric Toda
lattice equation \cite{DLS,KS1} after imposing the reduction
constrains $u_{0,i}=1$, $v_{0,2i+1}=0$.


 We call
equations \pp{LaxEq} for arbitrary $(K^+,K^-)$ the 2D fermionic
$(K^+,K^-)$-Toda lattice hierarchy.

\section{ 1D  fermionic $(K^+,K^-)$-Toda lattice
hierarchy}
 In this section, we consider the reduction of the 2D
fermionic $(K^+,K^-)$-Toda lattice hierarchy for even values of
$(K^+,K^-)$ to the 1D space.

Let $(K^+,K^-)$ be even numbers. In this case the generalized
graded bracket \pp{SK-bracket} between two $Z_2$-even operators
turns into the usual commutator and eqs. \pp{LaxEq} become
\begin{eqnarray}\label{LaxEqBos}
\ D_s^\pm L^\alpha_{K^\alpha}&=& [((L^\pm_{K^\pm})^s)_{\pm},
L^\alpha_{K^\alpha}].
 \eeaa
%
%

Following \cite{KaSa} for even $(K^+,K^-)$ one can impose the
reduction constraint
 on the Lax operators \pp{LaxOp} as follows:
\bbaa \label{redLax}
L^+_{K^+}+(L^+_{K^+})^{-1}=L^-_{K^-}+(L^-_{K^-})^{-1}\equiv
L_{K^+,K^-}, \eeaa
which  leads to the following explicit form for the reduced Lax
operator
\bbaa \label{LaxKM} L_{K^+,K^-}=\sum\limits_{k=0}^{K^+} u_{k,i}
e^{(K^+-k)\partial}+\sum\limits_{k=0}^{K^--1} v_{k,i}
e^{(k-K^-)\partial}\equiv\sum\limits_{k=0}^{K^++K^-} {\tilde
u}_{k,i} e^{(K^+-k)\partial}.
 \eeaa
Substituting the expressions for the reduced composite Lax
operators $L^\pm_{K^\pm}=L_{K^+,K^-}-(L^\pm_{K^\pm})^{-1}$ into
Lax equations \pp{LaxEqBos}  one can see that these equations
become equivalent to the single Lax equation on the reduced Lax
operator
\bbaa \label{1dLaxEq}D_sL_{K^+,K^-}&=&
[((L_{K^+,K^-})^s)_+,L_{K^+,K^-}] \eeaa
with $D_s^+=-D^-_s=D_s.$ As a consequence of eq. \pp{1dLaxEq} we
have
\bbaa \label{1dLaxEqCom}D_s(L_{K^+,K^-})^r&=&
[((L_{K^+,K^-})^s)_+, ( L_{K^+,K^-})^r].\nn \eeaa

At the reduction $u_{0,i}=1$ the 1D $(2,2)$-TL hierarchy becomes
that studied in detail in \cite{DDNS}. In this case, the
representation \pp{1dLaxEq}  with the Lax operator
\bbaa\label{LaxT}
L_{2,2}=e^{2\partial}+\gamma_ie^{\partial}+c_i+\rho_ie^{-\partial}+d_ie^{-2\partial}\nn
\eeaa
gives  the following first flow:
\bbaa\label{1dToda}
&&D_1 d_i=d_i(c_i-c_{i-2}),\ \ \nn\\
&&D_1 \rho_i=\rho_i(c_i-c_{i-1})+d_{i+1} \gamma_i-d_i\gamma_{i-2},
\nn \\
&&D_1\gamma_i=\rho_{i+2}-\rho_{i},\ \ \nn\\
 && D_1 c_i=d_{i+2}
-d_{i}+\gamma_i \rho_{i+1}+\gamma_{i-1}\rho_i. \nn\eeaa
These are the 1D generalized fermionic Toda lattice equations
\cite{DDNS}  which    possess the $N=4$ supersymmetry.

\section{Bi-Hamiltonian structure of 1D fermionic $(K^+,K^-)$-TL hierarchy}
 In this section, we apply the R-matrix
approach to build   the bi-Hamiltonian structure of the 1D
fermionic $(K^+,K^-)$-TL hierarchy and perform its Dirac
reduction.

The space of operators $\mathbb{O}_{K^+}^+$  \pp{O+-}
%
%
can obviously  be   split  into the vector space direct sum,
$\mathbb{O}_{K^+}=(\mathbb{O}_{K^+})_+\dotplus
(\mathbb{O}_{K^+})_-$.
The $R$-matrix arising from this splitting
\bbaa R=P_+-P_-,\ \ \ \ \ \
R(\mathbb{O}_{K^+})=(\mathbb{O}_{K^+})_+- (\mathbb{O}_{K^+})_- \nn
\eeaa
obviously solves the graded  modified  Yang-Baxter equation
\pp{YB} at $\alpha=1$. This $R$-matrix is not graded
antisymmetric, $R\neq -R^\dag$; however, its graded antisymmetric
part $A=1/2(R-R^\dag)$ as well as the $R$-matrix itself satisfy
the graded modified Yang-Baxter equation \pp{YB}. According to the
general Theorem of Section 3 this means that there exist two
Poisson structures on ${\sf g}^\dag={\sf g}.$ Substituting the
general form of operators $L_{K^+}^+$ \pp{LaxOp} and
\bbaa \label{grad1D} \nabla u_{n,\xi}=e^{(n-K^+)\partial}(-1)^i
\delta_{i,\xi} \nn\eeaa
into \pp{PB1} and \pp{PB2} one can find  the  explicit form of the
first and second Poisson brackets, respectively,
\bbaa \label{PB1d1} \{ u_{n,i},u_{m,j}
\}_1&=&(-1)^j(\delta^-_{n,K^+}+\delta^-_{m,K^+}-1) (u_{n+m-K^+,i}
\delta_{i,j+n-K^+}\nn
\\
&&-\ (-1)^{(m+K^+)(n+K^++1)}u_{n+m-K^+,j}, \delta_{i,j-m+K^+}
\eeaa
and
\bbaa
 \label{PB2d1}
 \{ u_{n,i},u_{m,j} \}_2 & =&\ -\ (-1)^j\frac12 \Bigr[ u_{n,i}
u_{m,j}( \delta_{i,j+n-K^+}
-(-1)^m   \delta_{i,j-m+K^+})    \nn  \\
&&+\ \sum\limits_{k=0}^{n+m} (\delta^+_{m,k}-\delta^-_{m,k})
\Bigr((-1)^{mk}u_{n+m-k,i} u_{k,j} \delta_{i,j+n-k}\nn\\
&& -\ (-1)^{m(n+k+1)}u_{k,i}
u_{n+m-k,j}\delta_{i,j-m+k}\Bigl)\Bigr],~~~~~~~ \eeaa
where
$$
\delta_{n,m}^+ =\left\{
   \begin{array}{rcl}
1,& \mbox{if} &n>m\\
0,& \mbox{if} &n\leq m \\
 \end{array}
\right. \ , \hspace*{1cm} \mbox{}
\delta_{n,m}^- =\left\{
   \begin{array}{rcl}
1,& \mbox{if} &n<m\\
0,& \mbox{if} &n\geq m. \\
 \end{array}
\right.
$$
Let us remind  that the second Poisson brackets are defined for
even values of $K^+$ only.

 Our next goal is to perform the reduction
of  Poisson brackets (\ref{PB1d1}--\ref{PB2d1})  for the functions
parameterizing the operators $L_{K^+}^+$ \pp{LaxOp} to the Poisson
brackets corresponding to the reduced operators \pp{LaxKM}
\bbaa\label{1dLax-red} L^{red}_{K^+,K^-}=e^{K^+\partial}+\sum_{
k=1 }^{K^++K^-}u_{k,i}e^{(K^+-k)\partial},\nn \eeaa
where $K^+, K^-$ are even numbers. Therefore, one needs to modify
the Poisson brackets (\ref{PB1d1}--\ref{PB2d1}) according to the
reduction constraints
\bbaa \label{redCon} \left. \begin{array}{rcl}
 u_{k,i}&=& 0,\ \ k>K^++K^-,\\
 u_{0,i}&=&1 \end{array} \right \} \mbox{ for any  } i.
\eeaa
We apply these reduction constraints in two steps. First, we note
that for the first constraint in \pp{redCon} the reduction simply
amounts to imposing   constraint $u_{k,i}= 0$ $( k>K^++K^-)$ due
to the observation that
\bbaa\label{pbCon1} \{u_{n,i},u_{m,j}\}_p \Biggr |_{\stackrel
{\scriptstyle u_{k,i}= 0,~}{\scriptscriptstyle k>K^++K^-}}
 =0,\ \ \ 0\leq n\leq
K^++K^-,\ \ m>K^++K^-,\ \ p=1,2. \eeaa
For the first Poisson brackets \pp{PB1d1} relation \pp{pbCon1} is
obvious. One can derive  eq. \pp{pbCon1} for the second Poisson
brackets \pp{PB2d1} if one divides the sum in \pp{PB2d1} into
three pieces
\bbaa \label{sumdiv}\sum\limits_{k=0}^{n+m}\ \  =
\sum\limits_{k=0}^{\mbox{\footnotesize max}(0,n+m-K^+-K^--1)}+
\sum\limits_{k=\mbox{\footnotesize
max}(1,n+m-K^+-K^-)}^{\mbox{\footnotesize min}(n+m-1,K^++K^-)}+
\sum\limits^{n+m}_{k=\mbox{\footnotesize min}(n+m,K^++K^-+1)}.
\eeaa
Now it is easy to verify  that the second sum in the r.h.s. of eq.
\pp{sumdiv} is the only sum which could  give a nonzero
contribution to eq. \pp{pbCon1}, but it is equal to zero if $0\leq
n\leq K^++K^-,\ \ m>K^++K^-.$

Now let us consider the second reduction constraint in eq.
\pp{redCon}, $u_{0,i}=1$.
  Following the
standard Dirac reduction prescription we obtain for the Dirac
brackets
\bbaa\label{PBred}
\{u_{n,i},u_{m,j}\}_p^{red}=\biggl(\{u_{n,i},u_{m,j}\}_p\biggl)\biggl|_{u_{0,i}=1}
-\ \triangle_p(u_{n,i},u_{m,j}), \ \ p=1,2 ~~~~~~~~\nn\eeaa
 with the correction term
\bbaa &&\triangle_p(u_{n,i},u_{m,j})=\Biggl(\sum_{i^{'},j^{'}}
\{u_{n,i},u_{0,i^{'}}\}_p
\{u_{0,i^{'}},u_{0,j^{'}}\}_p^{-1}\{u_{0,j^{'}},u_{m,j}\}_p\Biggl)\Biggl|_{u_{0,i}=1}.
\nn \eeaa

In the case of the first Poisson brackets
$\{u_{0,i},u_{m,j}\}_1=0$ for any $m$. Thus, one can conclude that
the first Poisson brackets \pp{PB1d1} are not modified, and they
can simply be restricted by imposing the constraints \pp{redCon}.

Before investigating the second Poisson brackets \pp{PB2d1} we
supply the  fields $u_{n,i}$ and $v_{n,i}$ with the boundary
conditions
\bbaa\label{bound} \lim_{i\rightarrow \pm
\infty}u_{n,i}=\lim_{i\rightarrow \pm \infty}v_{n,i}=0,\ \ \ \
n\neq 0 \eeaa
and introduce a new notation $\delta_{i,j-n}:=(\Lambda^n)_{i,j}$
which is useful in what follows. One can verify that in the new
notation the multiplication of matrices results in adding powers
of the operators $\Lambda$:
$$
\delta_{i,j^{'}+n}\delta_{j^{'},j+m}:=\Lambda^{-n}_{i,j^{'}}\Lambda_{j^{'},j}^{-m}=\Lambda_{i,j}^{-m-n}.
$$
Then, one can represent the correction term for the reduced second
Poisson brackets as follows:
 \bbaa
\triangle_2(u_{n,i},u_{m,j})&=&1/2(-1)^j
u_{n,i}[(1-\Lambda^{K^+-n})(1+\Lambda^{-K^+})\nn\\
&&(\Lambda^{K^+}-\Lambda^{-K^+})^{-1}(1+\Lambda^{K^+})(1-(-1)^m\Lambda^{m-K^+})]_{i,j}u_{m,j}.\nn
\eeaa
In general, the reduced second Poisson brackets are nonlocal since
the inverse matrix
$$(\Lambda^{2\nu}-\Lambda^{-2\nu})^{-1}=(1+\Lambda^{2\nu})^{-1}(1-\Lambda^{-\nu})^{-1}(1+\Lambda^{-\nu})^{-1},$$
 being considered as an operator acting in the space of
functionals with boundary conditions \pp{bound} can be  expressed
via infinite sums
\bbaa
(1-\Lambda^{-\nu})^{-1}&=&\lambda_1\sum_{k=0}^{\infty}\Lambda^{-k\nu}-(1-\lambda_1)\sum_{k=1}^{\infty}\Lambda^{k\nu},\nn\\
(1+\Lambda^{-\nu})^{-1}&=&\lambda_2\sum_{k=0}^{\infty}(-1)^k\Lambda^{-k\nu}+
(1-\lambda_2)\sum_{k=1}^{\infty}(-1)^k\Lambda^{k\nu}, \nn\eeaa
where $\lambda_1$ and $\lambda_2$ are arbitrary parameters.
However, in the particular case  $K^+=2$ the Dirac bracket becomes
local, since the nonlocality is eliminated due to the contraction
of the matrix with its  inverse matrix.  Indeed, one has
\bbaa
1-\Lambda^{\nu}&=&(1-\Lambda^{-1})(\delta^-_{\nu,0}\sum_{k=\nu+1}^{0}\Lambda^{k}-
\delta^+_{\nu,0}\sum_{k=1}^{\nu}\Lambda^{k}),\nn\\
1-(-1)^\nu\Lambda^{\nu}&=&
(1+\Lambda^{-1})(\delta^-_{\nu,0}\sum_{k=\nu+1}^{0}(-1)^k\Lambda^{k}-
\delta^+_{\nu,0}\sum_{k=1}^{\nu}(-1)^k\Lambda^{k})\nn\eeaa
and for $K^+=2$ the Dirac brackets are local
 \bbaa \label{PBred22}
\triangle_2(u_{n,i},u_{m,j})=1/2(-1)^j
u_{n,i}[(1+\Lambda^{-2})~~~~~~~~~~~~~~~~~~~~~~~~~~~~~~~~~~~~~~~~~~~~~~~~~\nn\\
\!\!\!
(\delta_{n,2}^+\sum_{k=3-n}^0\Lambda^k-\delta_{n,2}^-\sum_{k=1}^{2-n}\Lambda^k)
(\delta_{m,2}^-\sum_{s=m-1}^0(-1)^s\Lambda^s-\delta_{m,2}^+\sum_{s=1}^{m-2}(-1)^s\Lambda^s)]_{i,j}u_{m,j}.\eeaa

As an example, we finish this section discussing   the explicit
form of the second Hamiltonian structure  of the 1D $(2,2)$-Toda
lattice hierarchy. The Lax operator defining this hierarchy is
parameterized as follows:
$$ L_{2,2}=u_i
e^{2\partial}+\gamma_ie^{\partial}+c_i+\rho_ie^{-\partial}+d_ie^{-2\partial}.
$$
Using  eqs. \pp{PB1d1}, \pp{PB2d1} and \pp{PBred22} one can derive
the corresponding first Hamiltonian structure
\bbaa\label{2-2PB1}
\{d_i, c_j\}_1& =&  ( -1 ) ^j d_i( \delta_{i, j}  - \delta_{i, j+2 }),\nn \\
\{c_i,\rho_j\}_1& =&  -( -1 ) ^j \rho_j
        (\delta_{i, j - 1} +  \delta_{i, j}),\nn \\
  \{\rho_i,
      \rho_j\}_1 &=&      ( -1 ) ^j (d_i \delta_{i,
              j - 1}\  - d_j \delta_{i,j + 1})   ,\nn\\
  \{\gamma_i, \gamma_j\}_1& =&
     ( - 1 ) ^j ( \delta_{i, j - 1}- \delta_{i,j + 1})
\eeaa
and the second Hamiltonian structure
 \bbaa\label{2-2PB2}
\!\!\!\!\!\!\!\!\{d_i,d_j\}_2&=&1/2 (-1)^j
d_id_j(1+\Delta)(\delta_{i,j-2}-\delta_{i,j+2}),\nn\\
\!\!\!\!\!\!\!\!\{d_i,\rho_j\}_2&=&1/2 (-1)^jd_i\rho_j((1-\Delta)(\delta_{i,j}+\delta_{i,j+1})-(1+\Delta)(\delta_{i,j-1}+\delta_{i,j+2})),\nn\\
\!\!\!\!\!\!\!\!\{d_i,c_j\}_2&=&
(-1)^jd_ic_j(\delta_{i,j}-\delta_{i,j+2}),\nn\\
\!\!\!\!\!\!\!\!\{d_i,\gamma_j\}_2&=&1/2 (-1)^j d_i\gamma_j
((1-\Delta)(\delta_{i,j}+\delta_{i,j+3})-(1+\Delta)(\delta_{i,j+1}+\delta_{i,j+2})),\nn\\
\!\!\!\!\!\!\!\!\{d_i,u_j\}_2&=&1/2
(-1)^jd_iu_j(1-\Delta)(\delta_{i,j}-\delta_{i,j-4}),\nn\\
\!\!\!\!\!\!\!\!\{c_i,c_j\}_2&=&
(-1)^j(u_id_j\delta_{i,j-2}-u_jd_i\delta_{i,j+2}+\gamma_i\rho_j\delta_{i,j-1}+\gamma_j\rho_i\delta_{i,j+1}),\nn\\
\!\!\!\!\!\!\!\!\{c_i,\rho_j\}_2&=&
(-1)^j(d_i\gamma_j\delta_{i,j+1}+d_j\gamma_i\delta_{i,j-2}-
c_i\rho_j(\delta_{i,j}+\delta_{i,j-1})),\nn\\
\!\!\!\!\!\!\!\!\{c_i,\gamma_j\}_2&=&-
(-1)^j(u_i\rho_j\delta_{i,j-1}+u_j\rho_i\delta_{i,j+2}),\nn\\
\!\!\!\!\!\!\!\!\{\rho_i,\rho_j\}_2&=&
(-1)^j(c_id_j\delta_{i,j-1}-c_jd_i\delta_{i,j+1}-1/2\rho_i\rho_j
(1+\Delta)(\delta_{i,j+1}+\delta_{i,j-1})),\nn\\
\!\!\!\!\!\!\!\! \{\rho_i,\gamma_j\}_2&=&
(-1)^j(u_id_j\delta_{i,j-1}-u_jd_i\delta_{i,j+3}-
\rho_i\gamma_j(\delta_{i,j+1}-1/2(1-\Delta)(\delta_{i,j}+\delta_{i,j-2}))),\nn\\
\!\!\!\!\!\!\!\!\{\rho_i,u_j\}_2&=&1/2
(-1)^j\rho_iu_j(1-\Delta)(\delta_{i,j}-\delta_{i,j+1}+\delta_{i,j+2}-\delta_{i,j+3}),\nn\\
\!\!\!\!\!\!\!\!\{\gamma_i,\gamma_j\}_2&=& (-1)^j(u_i
c_j\delta_{i,j-1}-u_j
c_i\delta_{i,j+1}-1/2\gamma_i\gamma_j(1-\Delta)(\delta_{i,j+1}+\delta_{i,j-1})),\nn\\
\!\!\!\!\!\!\!\!\{\gamma_i,u_j\}_2&=&1/2 (-1)^j\gamma_iu_j(1-\Delta)(\delta_{i,j}-\delta_{i,j+1}+\delta_{i,j+2}-\delta_{i,j-1})),\nn\\
\!\!\!\!\!\!\!\!\{u_i,u_j\}_2&=&1/2
(-1)^ju_iu_j(1-\Delta)(\delta_{i,j+2}-\delta_{i,j-2}),
%
\eeaa
where only nonzero brackets are written down. Here we have
introduced the parameter $\Delta$ which for the unreduced brackets
is
 equal to zero, $\Delta=0$, and  for the Dirac reduced brackets
 with
 the reduction  constraint  $u_i=1$ is equal to one, $\Delta=1$.
  In the latter case,  algebras \pp{2-2PB1} and \pp{2-2PB2} reproduce, respectively, the
first and second  Hamiltonian structures of the 1D generalized
fermionic Toda
 lattice hierarchy found in \cite{DDNS} by a  heuristic approach.

\section{Bi-Hamiltonian structure of 2D fermionic $(K^+,K^-)$-TL hierarchy}
 In this section, we  construct the
bi-Hamiltonian structure of the 2D fermionic $(K^+,K^-)$-TL
hierarchy. This hierarchy is associated with two Lax operators
\pp{LaxOp} belonging to the operator space (\ref{O+-}). Following
\cite{Carlet} we consider the associative algebra on the space of
the direct sum of two
 difference operators
\bbaa \label{O2} {\sf g}:= \mathbb{O}^+_{K^+}\oplus
\mathbb{O}^-_{K^-}. \eeaa
However, in contrast  to the case of pure bosonic 2D TL hierarchy,
the difference operators in the direct sum \pp{O2} can be of both
opposite and  equal  diagonal $Z_2$-parity. It turns out that the
Poisson brackets can correctly be  defined only for the latter
case.  In what follows we restrict ourselves to the case when both
operators in {\sf g} \pp{O2} have the same diagonal parity.

%
%

 We denote  $(x^+,x^-)$  elements of such algebra ${\sf g}={\sf
g}^\dag$ with the product
\bbaa\label{product}
(x^+_1,x^-_1)\cdot(x^+_2,x^-_2)=(x^+_1x^+_2,x^-_1x^-_2), \eeaa
and define the inner product  as follows:
 \bbaa \label{Ipr2}
<(x^+, x^-)>:=str(x^++x^-), \eeaa
 where $x^+\in \mathbb{O}^+_{K_1}$, $x^-\in\mathbb{O}^-_{K_2}$.
 Using this definition we set the Poisson bracket
 to be
\bbaa \label{pb2D} \{f_1,f_2\}= <(\mathbb{O}^+_{K^+},
\mathbb{O}^-_{K^-}),[\nabla f_1,\nabla f_2\}^\oplus>, \eeaa
where $$[\nabla f_1,\nabla f_2\}^\oplus:= ([\nabla f_1^+,(\nabla
f_2^+)^{*(d_{\nabla f_1^+})}\},[\nabla f_1^-,(\nabla
f_2^-)^{*(d_{\nabla f_1^-})}\}),$$
$f_k$ are functionals on {\sf g} \pp{O2}, and $\nabla
f_k[(\mathbb{O}^+_{K^+},\mathbb{O}^-_{K^-})]=(\nabla f^+_k,\nabla
f^-_k)$ are their gradients which can be  found from the
definition
 \bbaa \frac{\partial f_k[(\mathbb{O}^+_{K^+},\mathbb{O}^-_{K^-})+\epsilon (\delta \mathbb{O}^+_{K^+},
\delta \mathbb{O}^-_{K^-})] }{\partial\epsilon}{\Biggr
|}_{\epsilon=0}\! &=&\! <(\delta \mathbb{O}^+_{K^+}, \delta
\mathbb{O}^-_{K^-}), (\nabla f^+_k,\nabla
f^-_k)>\nn\\
&\!=\! &<\delta \mathbb{O}^+_{K^+},\nabla f^+_k>+<\delta
\mathbb{O}^-_{K^-},\nabla f^-_k>.\nn \eeaa
 In order to
obtain nontrivial Hamiltonian dynamics,  one needs to modify the
bracket \pp{pb2D} applying the $R$-matrix
\bbaa [\nabla f_1,\nabla f_2\}^\oplus\ \ \longrightarrow \ \
[\nabla f_1,\nabla f_2\}^\oplus_R=[R(\nabla f_1),\nabla
f_2\}^\oplus+ [\nabla f_1,R(\nabla f_2)\}^\oplus. \nn\eeaa
 The $R$-matrix acts on the space \pp{O2} in the
nontrivial way  and mixes up the elements from two subalgebras in
the direct sum with each other
\bbaa \label{2dR} R(x^+,x^-)=(x^+_+-x^+_-+2x^-_-,x_-^-
-x^-_++2x^+_+) \eeaa
which is a crucial point of the $R$-matrix approach in the
two-dimensional case \cite{Carlet}. This $R$-matrix  allows one to
find two compatible Poisson structures and rewrite the Lax-pair
representation \pp{LaxEq} in the Hamiltonian form.

By construction the $R$-matrix \pp{2dR} satisfies the graded
modified
 Yang-Baxter equation
\bbaa \label{YB2D}R([(x^+,x^-),(y^+,y^-)\}_R)-
[R(x^+,x^-),R(y^+,y^-)\}=\alpha [(x^+,x^-),(y^+,y^-)\} \eeaa
with $\alpha=1$. In order to show this, we simply repeat the
arguments of \cite{Carlet}  representing the $R$-matrix as the
difference $R=\Pi-\bar\Pi$ of two projection operators
$$ \Pi(x^+,x^-)=(x^+_++x^-_-,x^+_++x^-_-), \ \ \ \ \ \
\bar\Pi(x^+,x^-)=(x^+_--x^-_-,x^-_+-x^+_+) $$
$$\Pi(x^+,x^-)+\bar\Pi(x^+,x^-)=(x^+,x^-),\ \ \ \ \Pi^2=\Pi,\ \ \ \
\bar\Pi^2=\bar\Pi,\ \ \ \ \bar\Pi\Pi=\Pi\bar\Pi=0.$$
 Therefore, the
 $R$-matrix \pp{2dR} provides the splitting of the algebra  {\sf g} and solves the modified graded
Yang-Baxter equation \pp{YB2D}.

The two-dimensional $R$-matrix is not graded antisymmetric, its
adjoint counterpart $R^\dag$ looks like
\bbaa \label{2dRs} R^\dag (x^+,x^-)=(x^+_{\leqslant 0}
-x^+_{>0}+2x^-_{\leqslant 0},x_{>0}^- -x^-_{\leqslant
0}+2x^+_{>0})=\Pi^\dag-\bar\Pi^\dag ,\nn\eeaa
where the dual projections are
\bbaa \Pi^\dag(x^+,x^-)&=&(x^+_{\leqslant 0}+x^-_{\leqslant 0}
,x^+_{>0}+x^-_{>0}), \ \ \ \ \nn\\
\bar\Pi^\dag(x^+,x^-)&=&(x^+_{>0}-x^-_{\leqslant 0},x^-_{\leqslant
0 }-x^+_{>0}). \nn\eeaa

The direct verification by substitution in \pp{YB2D} shows that
the graded antisymmetric part
\bbaa 1/2
(R(x^+,x^-)-R^\dag(x^+,x^-))=(x^+_{>0}-x^+_{<0}-x^-_0,x_{<0}^-
-x^-_{>0}+x^+_0) \nn\eeaa
also satisfies the   graded modified  Yang-Baxter equation
\pp{YB2D}. Therefore, by Theorem of Section~3 there exist two
Poisson structures on {\sf g} \pp{O2}.

Using  eqs. (\ref{PB1}--\ref{PB2}), (\ref{product}--\ref{Ipr2}),
\pp{2dR} and cyclic permutations inside the supertrace
\pp{supertr} we obtain the following general form of the first and
second Poisson brackets:
\bbaa\label{PB-2d} \{f,g\}_i&=& <P^+_i(\nabla g^+,\nabla
g^-),(\nabla f^+)^{*(d_{\nabla g})}>\nn\\
&+&<P^-_i(\nabla g^+,\nabla g^-),(\nabla f^-)^{*(d_{\nabla g})}>,\
\ \ \ \ i=1,2, \eeaa
where  $d_{\nabla g}:=d_{\nabla g^+}=d_{\nabla g^-}$.
 The Poisson tensors in eq. \pp{PB-2d}  are found for any values of $(K^+,K^-)$
 for the first Hamiltonian structure
\bbaa P_1^+(\nabla g^+,\nabla g^-) &=& [(\nabla g^-_-- \nabla
g^+_-)^{*(K^+)},(L^+_{K^+})^{*(d_{\nabla g})}\}
\nn\\
& -&([L^+_{K^+},\nabla g^+\}+[L^-_{K^-},\nabla g^-\})_{\leqslant 0},\nn\\
P_1^-(\nabla g^+,\nabla g^-) &=& [(\nabla g^+_+- \nabla
g^-_+)^{*(K^-)},(L^-_{K^-})^{*(d_{\nabla g})}\}
\nn\\
&-&([L^+_{K^+},\nabla g^+\}+[L^-_{K^-},\nabla g^-\})_{>
0},\nn\eeaa
while for the second Hamiltonian structure we constructed the
explicit   expression of the Poisson tensors  for even values of
$(K^+,K^-)$ only
\bbaa P_2^+(\nabla g^+,\nabla g^-) &=&\ 1/2\Bigr ([(\nabla
g^-(L^-_{K^-})^{*(d_g)}+L^-_{K^-}\nabla g^-
\nn\\
&-&\nabla g^+(L^+_{K^+})^{*(d_g)}
-L^+_{K^+}\nabla g^+)_-,(L^+_{K^+})^{*(d_g)}\}\nn\\
 & -& L^+_{K^+}\ ([L^+_{K^+},\nabla g^+\}+[L^-_{K^-},\nabla
g^-\})_{\leqslant 0}
\nn\\
&-&([L^+_{K^+},\nabla g^+\}+[L^-_{K^-},\nabla g^-\})_{\leqslant 0}\ (L^+_{K^+})^{*(d_g)}\Bigl),\nn\\
P_2^-(\nabla g^+,\nabla g^-)&=&1/2\Bigr ([(\nabla
g^+(L^+_{K^+})^{*(d_g)}+L^+_{K^+}\nabla g^+
\nn\\
&-&\nabla g^-(L^-_{K^-})^{*(d_g)}
-L^-_{K^-}\nabla g^-)_+,(L^-_{K^-})^{*(d_g)}\}\nn\\
&-&L^-_{K^-}\ ([L^+_{K^+},\nabla g^+\}+[L^-_{K^-},\nabla
g^-\})_{>0}
\nn\\
&-&([L^+_{K^+},\nabla g^+\} +[L^-_{K^-},\nabla g^-\})_{> 0}\
(L^-_{K^-})^{*(d_g)}\Bigl).\nn
 \eeaa
The Poisson brackets for the functions $u_{n,i}$ and $v_{n,i}$
parameterizing the Lax operators \pp{LaxOp} can  explicitly  be
derived from \pp{PB-2d} if one takes into account  that
\bbaa\label{grad-uv} \nabla u_{n,\xi}&\equiv&(\nabla
u_{n,\xi}^+,\nabla u_{n,\xi}^-)=(e^{(n-K^+)\partial}(-1)^i
\delta_{i,\xi},0),\nn\\
 \nabla v_{n,\xi}&\equiv&(\nabla v_{n,\xi}^+,\nabla
v_{n,\xi}^-)=(0,e^{(K^--n)\partial}(-1)^i \delta_{i,\xi}).\nn\eeaa
In such  a way one can obtain the following expressions:
\bbaa\label{PB1-func} \{
u_{n,i},u_{m,j}\}_1&=&(-1)^j(\delta^-_{n,K^+}+\delta^-_{m,K^+}-1)
\Big(u_{n+m-K^+,i} \delta_{i,j+n-K^+}\nn\\
&&-\ (-1)^{(m+K^+)(n+K^++1)}u_{n+m-K^+,j} \delta_{i,j-m+K^+}\Big),\nn\\
\{ u_{n,i},v_{m,j} \}_1&=&(-1)^j\biggl[ \delta^+_{m,K^-}
\Big((-1)^{(m+K^-)(n+K^++1)}u_{n-m+K^-,j}
\delta_{i,j+m-K^-}\nn\\
&& -\ u_{n-m+K^-,i} \delta_{i,j+n-K^+}\Big) + (\delta^-_{n,K^+}-1)
\Big(v_{m-n+K^+,i}
\delta_{i,j+n-K^+}\nn\\
&&-\ (-1)^{(m+K^-)(n+K^++1)}v_{m-n+K^+,j} \delta_{i,j+m-K^-}\Big)
\biggr],\nn\\
\{
v_{n,i},v_{m,j}\}_1&=&(-1)^j(1-\delta^+_{n,K^-}-\delta^+_{m,K^-})
\Big(v_{n+m-K^-,i} \delta_{i,j-n+K^-}\nn\\
&&-\ (-1)^{(m+K^-)(n+K^-+1)}u_{n+m-K^-,j}\delta_{i,j+m-K^-}\Big)
\eeaa
for the first Hamiltonian structure and
%
\bbaa\label{PB2-func} \{ u_{n,i},u_{m,j} \}_2&=&-\ (-1)^j\frac12
\Bigr[ u_{n,i} u_{m,j}( \delta_{i,j+n-K^+}
-(-1)^m   \delta_{i,j-m+K^+})  \nn  \\
&&+\ \sum\limits_{k=0}^{n+m} (\delta^+_{m,k}-\delta^-_{m,k})
\Bigr((-1)^{mk}u_{n+m-k,i} u_{k,j} \delta_{i,j+n-k}
   \nn \\
   &&
-\ (-1)^{m(n+k+1)}u_{k,i} u_{n+m-k,j}\delta_{i,j-m+k}\Bigl)\Bigr],\nn\\
\{ u_{n,i},v_{m,j} \}_2&=&-(-1)^j\frac12
 \Bigr[ u_{n,i}  v_{m,j}\Big( \delta_{i,j}+
   \delta_{i,j+n-K^+}\nn\\
&&-\ (-1)^m (  \delta_{i,j+m-K^-} + \delta_{i,j+n+m-K^+-K^-})\Big)
  \nn \\
&&+\ 2\sum\limits_{k=\mbox{\footnotesize max}(0,m-n)}^{m-1}
\Bigr(u_{n-m+k,i} v_{k,j+m-k} \delta_{i,j+n-K^+}\nn\\
&&-\ (-1)^{(n+1)m} v_{k,j} u_{n-m+k,i+k-m}
\delta_{i,j+m+K^-}\Bigl)\Bigr], \nn\\
\{ v_{n,i},v_{m,j} \}_2&=&(-1)^j\frac12
 \Bigr[ v_{n,i} v_{m,j}( \delta_{i,j-n+K^-}
-(-1)^m   \delta_{i,j+m-K^-})
  \nn\\
&&-\ \sum\limits_{k=0}^{n+m} (\delta^+_{m,k}-\delta^-_{m,k})
\Bigr((-1)^{mk}v_{n+m-k,i} v_{k,j} \delta_{i,j-n+k} \nn\\
&&-\ (-1)^{m(n+k+1)}v_{k,i} v_{n+m-k,j}
\delta_{i,j+m-k}\Bigl)\Bigr] \eeaa
for the second Hamiltonian structure; the latter  is valid for
even values of $(K^+,K^-)$ only.

The reduction, according the reduction constraint $u_{0,i}=1$,
does not require any correction terms for the first Hamiltonian
structure \pp{PB1-func} because $\{u_{0,i},u_{n,j}
\}=\{u_{0,i},v_{n,j} \}=0$. For the second Hamiltonian structure
the correction terms are
 \bbaa \label{cor-2d}\triangle_2(u_{n,i},u_{m,j})&=&1/2(-1)^j
u_{n,i}(1-\Lambda^{K^+-n})(1+\Lambda^{-K^+})\nn\\
&&(\Lambda^{K^+}-\Lambda^{-K^+})^{-1}(1+\Lambda^{K^+})(1-(-1)^m\Lambda^{m-K^+})u_{m,j},\nn\\
\triangle_2(u_{n,i},v_{m,j})&=&1/2(-1)^j
u_{n,i}(1-\Lambda^{K^+-n})(1+\Lambda^{-K^+})\nn\\
&&(\Lambda^{K^+}-\Lambda^{-K^+})^{-1}(1+\Lambda^{K^+})(1-(-1)^m\Lambda^{K^--m})v_{m,j},\nn\\
\triangle_2(v_{n,i},v_{m,j})&=&1/2(-1)^j
v_{n,i}(1-\Lambda^{n-K^-})(1+\Lambda^{-K^+})\nn\\
&&(\Lambda^{K^+}-\Lambda^{-K^+})^{-1}(1+\Lambda^{K^+})(1-(-1)^m\Lambda^{K^--m})v_{m,j}
\eeaa
and they are nonlocal due to the presence of
$(\Lambda^{K^+}-\Lambda^{-K^+})^{-1}$ in the r.h.s.  of  eq.
\pp{cor-2d}. However, there are unique values of $(K^+,K^-)$ when
nonlocal terms are eliminated. Indeed, for
 $K^+=K^-=2$ eqs. \pp{cor-2d} become local
 \bbaa
&&\triangle_2(u_{n,i},u_{m,j})\ =\ 1/2(-1)^j
u_{n,i}(1+\Lambda^{-2})\nn\\
&&~~~~~~(\delta_{n,2}^+\sum_{k=3-n}^0\Lambda^k-\delta_{n,2}^-\sum_{k=1}^{2-n}\Lambda^k)
(\delta_{m,2}^-\sum_{s=m-1}^0(-1)^s\Lambda^s-\delta_{m,2}^+\sum_{s=1}^{m-2}(-1)^s\Lambda^s)u_{m,j},\nn\\
&&\triangle_2(u_{n,i},v_{m,j})\ =\ 1/2(-1)^j
u_{n,i}(1+\Lambda^{-2})\nn\\
&&~~~~~~(\delta_{n,2}^+\sum_{k=3-n}^0\Lambda^k-\delta_{n,2}^-\sum_{k=1}^{2-n}\Lambda^k)
(\delta_{m,2}^+\sum_{s=3-m}^0(-1)^s\Lambda^s-\delta_{m,2}^-\sum_{s=1}^{2-m}(-1)^s\Lambda^s)v_{m,j},\nn\\
&&\triangle_2(v_{n,i},v_{m,j})\ =\ 1/2(-1)^j
v_{n,i}(1+\Lambda^{-2})\nn\\
&&~~~~~~(\delta_{n,2}^-\sum_{k=n-1}^0\Lambda^k-\delta_{n,2}^+\sum_{k=1}^{n-2}\Lambda^k)
(\delta_{m,2}^+\sum_{s=3-m}^0(-1)^s\Lambda^s-\delta_{m,2}^-\sum_{s=1}^{2-m}(-1)^s\Lambda^s)v_{m,j}.\nn
\eeaa

The Hamiltonian structures thus obtained possess  the properties
(\ref{sym}--\ref{Jac}) with
$d_\mathbb{{O}}=d_{L^+_{K^+}}=d_{L^-_{K^-}}$. Using them one can
rewrite flows (\ref{Eqs1}--\ref{Eqs4}) for even values of
$(K^+,K^-)$ in the bi-Hamiltonian form
\bbaa D^\pm_s \Biggr( \begin{array}{c} u_{n,i}^{(r)}\\
v_{n,i}^{(r)}\end{array}\Biggr) =\{\Biggr( \begin{array}{c} u_{n,i}^{(r)}\\
v_{n,i}^{(r)}\end{array}\Biggr),H^\pm_{s+1}\}_1=\{\Biggr( \begin{array}{c} u_{n,i}^{(r)}\\
v_{n,i}^{(r)}\end{array}\Biggr),H^\pm_{s}\}_2,\nn\ \ \ \ \eeaa
with  Hamiltonians
\bbaa \label{ham2D}
H^+_s&=&\frac1sstr(L^+_{K^+})^s_*=\frac1s\sum_{i=-\infty}^\infty
(-1)^iu_{sK^+,i}^{(s)},\nn\\ \ \ \ \ \ \ \
H^-_s&=&\frac1sstr(L^-_{K^-})^s_*=\frac1s\sum_{i=-\infty}^\infty
(-1)^iv_{sK^-,i}^{(s)}.\eeaa
For odd values  of $(K^+,K^-)$ one can reproduce  the bosonic
flows of (\ref{Eqs1}--\ref{Eqs4}) only. In this case eqs.
\pp{ham2D}, due to relation \pp{evenH}, give only fermionic
nonzero Hamiltonians using which the bosonic flows can be
generated via odd first Hamiltonian structure \pp{PB1-func}
\begin{eqnarray} D^\pm_{2s} \Biggr( \begin{array}{c} u_{n,i}^{(r)}\\
v_{n,i}^{(r)}\end{array}\Biggr) =\{\Biggr( \begin{array}{c} u_{n,i}^{(r)}\\
v_{n,i}^{(r)}\end{array}\Biggr),H^\pm_{2s+1}\}_1.\nn\ \ \ \ \eeaa

One remark is in order. In Sec. 6, we derived the bi-Hamiltonian
structure for the 1D fermionic $(K^+,K^-)$-TL hierarchy in the
$R$-matrix approach applying the $R$-matrix formalism developed in
Sec. 3. However, the 1D fermionic $(K^+,K^-)$-TL hierarchy is
obtained as a reduction of the 2D fermionic $(K^+,K^-)$-TL
hierarchy with the reduction constraint \pp{redLax}. Therefore,
this reduction constraint can be carried over into Hamiltonian
structures and the bi-Hamiltonian structure for the 1D fermionic
$(K^+,K^-)$-TL hierarchy can equivalently be derived from  that of
the 2D hierarchy just by reduction with the corresponding
constraint. Actually,
this reduction amounts to the extraction of  subalgebras in the
Hamiltonian structures for the 2D fermionic $(K^+,K^-)$-TL
hierarchy. Indeed, one can verify that the fields $u_{n,i}\ (0\leq
n \leq K^+)$ and $v_{n,i}\ (0\leq n \leq K^--1)$ form subalgebras
for even values of $(K^+,K^-)$ in both the first \pp{PB1-func} and
the second \pp{PB2-func} Hamiltonian structures and these
subalgebras are  the first \pp{PB1d1} and the second \pp{PB2d1}
Hamiltonian structures, respectively, if one redefines the fields
as follows:
\bbaa u_{n,i}&=&\ u_{n,i},\ \ \ \ 0\leq n \leq K^+ \nn\\
u_{K^++K^--n,i}&=&-v_{n,i},\ \  0\leq n \leq K^--1. \nn \eeaa

\section{Conclusion}

 In this paper, we have generalized the $R$-matrix method
to the case of $Z_2$-graded operators with an involution and found
that there exist two Poisson bracket structures.  The first
Poisson bracket is defined for both odd and even operators with
$Z_2$-grading while the second one is found for even operators
only. It was shown that properties of the Poisson brackets were
provided by the properties of the generalized graded bracket. We
have deduced the operator form of the graded modified Yang-Baxter
equation and demonstrated that for the class of graded
antisymmetric $R$-matrices it was equivalent to the tensor form of
the graded classical Yang-Baxter equation.
 Then we have proposed the Lax-pair representation in terms of the
generalized graded bracket of the new 2D fermionic
$(K^+,K^-)$-Toda lattice hierarchy and demonstrated that this
hierarchy included all known up to now 2D supersymmetric TL
equations as subsystems. Next we have considered the reduction of
this hierarchy to the 1D space and reproduced  the 1D generalized
fermionic TL equations \cite{DDNS}. Finally, we  have applied the
developed R-matrix formalism to derive the bi-Hamiltonian
srtucture of the 1D and 2D fermionic $(K^+,K^-)$-TL hierarchies.
For even values of $(K^+,K^-)$ both even first and second
Hamiltonian structures were obtained and for this case all the
flows of the 2D fermionic $(K^+,K^-)$-TL hierarchy can be
rewritten in
 a bi-Hamiltonian form. For odd values of $(K^+,K^-)$
odd first Hamiltonian structure was found and  for this case only
bosonic flows of the 2D fermionic $(K^+,K^-)$-TL hierarchy can be
represented in a Hamiltonian form using fermionic Hamiltonians.

Thus, the problem of Hamiltonian description of the fermionic
flows of the 2D fermionic $(K^+,K^-)$-TL hierarchy  is still open.
Other problems  yet  to be answered are the construction of the
second Hamiltonian structure (if any) for odd Lax operators and of
the Hamiltonian structures (if any) for Lax operators $L_K^+$ and
$L_M^-$ of opposite $Z_2$-parities. All these questions are a
subject for future investigations.


\vspace{.8cm}
 {\bf Acknowledgments.  } We would like to thank A.P. Isaev, P.P.
 Kulish, and  A.A. Vladimirov
 for useful discussions.  This work was
partially supported by  RFBR-DFG Grant No. 04-02-04002, DFG Grant
436 RUS 113/669-2, the NATO Grant PST.GLG.980302, and by the
Heisenberg-Landau program.

\vspace{.3cm}

\vspace{.8cm}

 {\bf \Large Appendix A}

 \renewcommand{\theequation}{A.\arabic{equation}}
 \setcounter{equation}{0}

\vspace{.3cm}

Here we show that the graded modified Yang-Baxter equation \pp{YB}
for the case of  graded antisymmetric operators $R$ is equivalent
to the tensor form of the graded classical Yang-Baxter equation
introduced in the pioneer paper \cite{KulSk}
\bbaa \label{YB-op} [ r_{12},  r_{13}+ r_{23}]+[ r_{13},r_{23}]=0.
\eeaa

 Let $\mathcal{G}$ be a superalgebra with the generators
$e_\mu$ $(\mu =1, \dots,n+m)$, the structure constants $C_{\mu\nu
}^\rho $ and the graded Lie bracket
\bbaa \label{Lie-br} \{e_\mu,e_\nu\}=e_\mu e_\nu-(-1)^{d_\mu
d_\nu}e_\nu e_\mu=C_{\mu\nu}^\rho e_\rho, \nn\eeaa
where $ d_\mu$ is the Grassmann parity of the generator $e_\mu$
and $d_\mu=0$, if $\mu =1,\dots, n$ and $d_\mu=1$, if $\mu
=n+1,\dots, n+m$.

For the generators of the algebra $\mathcal{G}$ the graded
modified Yang-Baxter equation \pp{YB} at $\alpha=1$ takes the form
\bbaa  R(\{ R\, e_\xi,e_\gamma\})+R(\{ e_\xi,R\, e_\gamma\})-\{
R\, e_\xi,R\, e_\gamma \}=\{ e_\xi,e_\gamma\} \nn\eeaa
which can equivalently be rewritten as follows:
 \bbaa \label{YBcomp}(
R_\xi^\beta C_{\beta\gamma}^\alpha R_\alpha ^\mu +R_\gamma^\beta
C_{\xi\beta}^\alpha R_\alpha^\mu -R_\gamma^\beta R_\xi^\alpha
C_{\alpha \beta}^\mu)\, e_\mu = C_{\xi\gamma}^\mu e_\mu. \eeaa

Let us define an invariant supersymmetric non-degenerate bi-linear
form associated with some representation of $\mathcal{G}$
\bbaa <x,y>:=str(x\, y) \ \ \ \ \ \mbox{     for any } x,y \in
\mathcal{G} \nn\eeaa
using which one can introduce the dual basis $e^\mu$ in
superalgebra $\mathcal{G}$
\bbaa <e^\mu,e_\nu>=\delta^\mu_\nu \nn\eeaa
and the supermetric
\bbaa \eta_{\mu\nu}=<e_\mu,e_\nu>, \ \ \
\eta^{\mu\nu}=<e^\mu,e^\nu>, \ \ \ \eta^{\mu\nu}=-(-1)^{d_\mu}
\eta^{\nu\mu}=(\eta_{\nu\mu})^{-1} \nn\eeaa
by which one can raise and lower indices as follows:
\bbaa \label{up-down}
 e_\alpha=\eta_{\alpha\beta}e^\beta, \ \ \
e^\alpha=(-1)^{d_\alpha} \eta^{\alpha\beta}e_\beta=e_\beta
\eta^{\beta\alpha}, \ \ \
\eta_{\alpha\nu}C_{\gamma\mu}^\nu\eta^{\beta\mu}=C_{\alpha\gamma}^\beta.
\eeaa
Note that supermetric $\eta_{\mu\nu}$ and $R^\mu_\nu$ are  even
matrices, i.e., for any their nonzero entry one has $d_\mu +d_\nu
=0$.

Now we take the graded tensor product of both sides of eq.
\pp{YBcomp} with $e^\xi\otimes e^\gamma $ and using relations
\pp{up-down} rewrite it in the following form:
\bbaa \label{YBcomp2} ( (-1)^{d_\lambda}R^{\nu\beta} R^{\alpha
\mu} C_{\alpha\beta}^\lambda +(-1)^{d_\mu} R^{\lambda\beta}
R^{\alpha\mu}C_{\beta\alpha}^\nu &&\nn\\
-(-1)^{d_\mu}R^{\lambda\beta} R^{\nu\alpha} C_{\alpha \beta}^\mu
)\, e_\mu \otimes e_\nu\otimes e_\lambda&=&  \eta^{\nu\xi}
\eta^{\lambda\gamma} C_{\xi\gamma}^\mu \, e_\mu\otimes
e_\nu\otimes e_\lambda,
 \eeaa
 where
\bbaa
R^{\alpha\beta}=(-1)^{d_\alpha}\eta^{\alpha\gamma}R_\gamma^\beta,
\ \ \ R_\alpha^\beta=\eta_{\alpha\gamma}R^{\gamma\beta}.\nn\eeaa
 Here we use the graded tensor product
 \bbaa \label{ten-prod}
 (e_\mu\otimes e_\nu) (e_\lambda\otimes e_\xi)=(-1)^{d_\nu
 d_\lambda}(e_\mu e_\lambda \otimes e_\nu e_\xi),\\
\label{ten-prod1} (e_\mu\otimes e_\nu)^{i_1\, i_2}_{j_1\,
j_2}=(e_\mu)_{i_1 j_1}(e_\nu )_{i_2 j_2}
(-1)^{d_{i_2}(d_{i_1}+d_{j_1})}, \eeaa
where $d_i (d_j)$ means the Grassmann parity of the row (column)
of the supermatrix element $(e_\mu)_{ij}$ \cite{KulSk} and one has
$d_i+d_j=d_\mu$ for any nonzero $(e_\mu)_{ij}$.

Assume further that the operator $R$ is graded antisymmetric,
i.e.,
\bbaa R^{\mu\nu}=-(-1)^{d_\mu} R^{\nu\mu}\ \ \Longleftrightarrow \
\ < R\, x,y>=-<x,R\,y> \nn\eeaa
and define two $(n+m)\times (n+m)$ matrices
 \bbaa \tilde r =R^{\alpha\beta}e_\alpha\otimes e_\beta,\ \ \ \ t =(-1)^{d_\alpha}\eta^{\alpha\beta}e_\alpha\otimes
 e_\beta,\nn
 \eeaa
where $\tilde r$ is antisymmetric, $\tilde r^{i_1\, i_2}_{j_1\,
j_2}=-\tilde r^{i_2\, i_1}_{j_2\, j_1}$ and $t$ is the tensor
Casimir element invariant with respect to the adjoint action
 \bbaa  [t,x\otimes 1+1\otimes x]=0 \ \ \ \ \ \mbox{     for any } x \in
\mathcal{G}.\nn \eeaa
Note that $\tilde r$ and $t$ are even matrices, i.e.,
$d_{i_1}+d_{i_2}+d_{j_1}+d_{j_2}=0$ for any $\tilde r^{i_1\,
i_2}_{j_1\, j_2}\neq 0$ or $t^{i_1\, i_2}_{j_1\, j_2}\neq 0$.

Defining triple graded tensor products
\bbaa && \tilde r_{12} =R^{\alpha\beta}(e_\alpha\otimes e_\beta
\otimes 1), \ \ \ \tilde r_{13} =R^{\alpha\beta}(e_\alpha\otimes 1
\otimes e_\beta), \ \ \ \tilde r_{23} =R^{\alpha\beta} (1\otimes
e_\alpha
\otimes e_\beta),\nn\\
&& t_{12} =\eta^{\beta\alpha}(e_\alpha\otimes e_\beta \otimes 1),
\ \ \ \ t_{13} =\eta^{\beta\alpha}(e_\alpha\otimes 1 \otimes
e_\beta), \ \ \ \  t_{23} =\eta^{\beta\alpha} (1\otimes e_\alpha
\otimes e_\beta)\nn
 \eeaa
and using eqs. \pp{Lie-br} and \pp{ten-prod} one can find that eq.
\pp{YBcomp2} reads
\bbaa \label{YB-op-mod}[\tilde r_{12},\tilde  r_{13}+\tilde
r_{23}]+[\tilde r_{13},\tilde r_{23}]=-[t_{12},t_{13}] \eeaa
that is the tensor form of the graded modified classical
Yang-Baxter equation (see e.g. \cite{Isaev}). In order to
reproduce the graded classical Yang-Baxter equation with the zero
in the r.h.s one needs to introduce a new matrix
\bbaa r=\tilde r+t,\ \ \ \tilde r_{12}=1/2 (r_{12}-r_{21}),\ \ \
t_{12}=1/2 (r_{12}+r_{21}) \nn\eeaa
for which eq. \pp{YB-op-mod} takes the  form \pp{YB-op}.
Thus, for the case of  graded antisymmetric  operators $R$ we have
established the equivalence of equations \pp{YB} at $\alpha=1$ and
\pp{YB-op} which are, respectively, the operator form of the
graded modified classical Yang-Baxter equation and the tensor form
of the graded classical Yang-Baxter equation. Note that the former
equation is a more general one, since it admits solutions which
are not graded antisymmetric.

 For completeness we give here the component form of eq. \pp{YB-op}
 \bbaa && r^{i_1 i_2}_{k\,\, j_2} r^{k \,\, i_3}_{j_1
j_3} (-1)^{d_{j_2} (d_{i_3}+d_{j_3})} -  r^{i_1 i_3}_{k\,\, j_3}
r^{k\,\, i_2}_{j_1 j_2} (-1)^{d_{i_2} (d_{i_3}+d_{j_3})}\nn\\
&+& r^{i_1 i_3}_{ j_1 \, k} r^{i_2 \, k}_{j_2 j_3} (-1)^{d_{i_2}
(d_{i_1}+d_{j_1})} -  r^{i_2 i_3}_{ j_2\, k } r^{ i_1\, k}_{j_1
j_3} (-1)^{d_{j_2}
(d_{i_1}+d_{j_1})}\nn\\
&+& r^{i_1 i_2}_{ j_1\, k} r^{k\,\, i_3}_{j_2 j_3}  -
 r^{i_2 i_3}_{k\,\, j_3} r^{ i_1\, k}_{j_1 j_2} =0,\nn
 \eeaa
where eq. \pp{ten-prod1} is used when calculating the triple
tensor products.

\vspace{.3cm}

\vspace{.8cm}


 {\bf \Large Appendix B}

 \renewcommand{\theequation}{B.\arabic{equation}}
 \setcounter{equation}{0}

\vspace{.3cm}

 In this Appendix we investigate  bi-linear bracket \pp{PB2}
and establish  conditions on the $R$-matrix and its graded
antisymmetric part which are necessary in order the bracket
\pp{PB2}  be the Poisson bracket. Therefore, we need to verify the
Jacobi identities for any $f, g $ and $h$ in $\textit{\sf g}$
\bbaa \label{Jac1}   (-1)^{d_h d_g}\{h,\{f,g\}_2\}_2&+&\mbox{\rm
c. p.}\ \ =\ \
 -1/4(-1)^{d_h d_g}< [\mathbb{O},\nabla
(\{f,g\}_2)\}\!\cdot \nn\\
&&R\Big((\nabla h)^{*(d_g+d_f)} \mathbb{O}^{*(d_f+d_g+d_h)}
+\mathbb{O}^{*(d_g+d_f)}(\nabla h)^{*(d_g+d_f)}\Big)\nn\\
&&-\ R\Big(\nabla (\{f,g\}_2)
\mathbb{O}^{*(d_g+d_f)}+\mathbb{O}\nabla
(\{f,g\}_2)\Big)\!\cdot \nn\\
&&[\mathbb{O}^{*(d_g+d_f)},(\nabla h)^{*(d_g+d_f)}\}>\ \ +\ \
\mbox{\rm c. p.}\ =\ 0,
\eeaa
where
\bbaa\label{nablaPB}
 \nabla (\{f,g\}_2)&=&-1/4\biggr[\nabla g \
R((\nabla f)^{*(d_g)}
\mathbb{O}^{*(d_f+d_g)}+\mathbb{O}^{*(d_g)}(\nabla
f)^{*(d_g)})\nn\\
&&+ R(\mathbb{O}\nabla g+\nabla g\mathbb{O}^{*(d_g)})\ (\nabla
f)^{*(d_g)})\nn\\
&&+\nabla g \ R^\dag((\nabla
f)^{*(d_g)}\mathbb{O}^{*(d_f+d_g)}-\mathbb{O}^{*(d_g)}(\nabla
f)^{*(d_g)})\nn\\
&&+ R^\dag(\mathbb{O}\nabla g-\nabla g\mathbb{O}^{*(d_g)})\
(\nabla
f)^{*(d_g)})\nn\\
&&-(-1)^{d_f d_g}\biggl (R(\nabla
f\mathbb{O}^{*(d_f)}+\mathbb{O}\nabla f)\ (\nabla g)^{*(d_f)}\nn\\
&& + \nabla f \ R((\nabla g)^{*(d_f)}
\mathbb{O}^{*(d_f+d_g)}+\mathbb{O}^{*(d_f)}(\nabla
g)^{*(d_f)})\nn\\
&&+ \nabla f \ R^\dag((\nabla g)^{*(d_f)}
\mathbb{O}^{*(d_f+d_g)}-\mathbb{O}^{*(d_f)}(\nabla
g)^{*(d_f)})\nn\\
&&+R^\dag(\nabla f\mathbb{O}^{*(d_f)}-\mathbb{O}\nabla f)\ (\nabla
g)^{*(d_f)}
 \biggr) \biggr]. \eeaa
Inserting \pp{nablaPB} into \pp{Jac1}  after tedious but
straightforward calculations we get for the Jacobi identities
\bbaa \!\!\!\!\!\! (-1)^{d_f (d_h+d_g )}\! < [\mathbb{O},h\}\Big (
[R^\dag F_-,R^\dag G_- \} +[RF_+,RG_+\}-R([F_+,G_+\}_R)\Big)\!
>+\mbox{c.p.}=0,\nn
 \eeaa
where  we introduce the notation
\bbaa F_\pm&=&\mathbb{O}^{*(d_h)}(\nabla f)^{*(d_h)}\pm (\nabla
f)^{*(d_h)} \mathbb{O}^{*(d_h+d_f)}\nn\\
G_\pm&=&\mathbb{O}^{*(d_h+d_f)}(\nabla g)^{*(d_h+d_f)}\pm (\nabla
g)^{*(d_h+d_f)} \mathbb{O}^{*(d_h+d_f+d_g)}\nn
 \eeaa
for  brevity. For any linear map $R$ and its graded antisymmetric
part $A=1/2(R-R^\dag)$ one has the identity
\bbaa \label{ind} && (-1)^{d_f (d_h+d_g )}<[\mathbb{O},h\}[R^\dag
F_-,R^\dag G_-
\} >+\mbox{c.p.}\nn\\
&&~~~~~~~~~~~~=
(-1)^{d_f (d_h+d_g )}<[\mathbb{O},h\}\biggl ( 4/3
([AF_-,AG_-\}-A([F_+,G_+\}_A)\nn\\
&&~~~~~~~~~~~~~~~ -([RF_-,RG_-\}-R([F_-,G_-\}_R)\biggl
)>+\mbox{c.p.} \eeaa
 Now using \pp{ind} and the following identity
\bbaa (-1)^{d_f (d_h+d_g )}<[\mathbb{O},h\}\biggl ( [F_-,G_- \} +3
[F_+,G_+\})\biggl )>+\mbox{c.p.}=0,\nn \eeaa
which can directly be  verified,  we  finally rewrite the Jacobi
identities with an arbitrary parameter $\alpha$ as follows:
\bbaa && (-1)^{d_f (d_h+d_g )}<[\mathbb{O},h\}\biggl ( [RF_+,RG_+
\} -R([F_+,G_+\}_R)+\alpha [F_+,G_+\}\nn\\
&&~~~~~~~~~~~~~~~-([RF_-,RG_-
\} -R([F_-,G_-\}_R)+\alpha [F_-,G_-\}\nn\\
&&~~~~~~~~~~~~~~~+4/3([AF_-,AG_- \} -A([F_-,G_-\}_A)+\alpha
[F_-,G_-\} )\biggl )>+\mbox{c.p.}=0.\nn \eeaa
Now it is obvious that the Jacobi identities are satisfied if $R$
and its graded antisymmetric part $A$ obey the  graded  modified
Yang-Baxter equation \pp{YB} with the same $\alpha$.


\begin{thebibliography}{99}
\bibitem{Mikh} A.V. Mikhailov, Pisma Zh. Eksp. Teor. Fiz. {\bf 30}
(1979) 443.


\bibitem{UT} K. Ueno and K. Takasaki, "Toda Lattice Hierarchy",
{\it Adv. Stud. in Pure Math.}  {\bf 4} (1984) 1

\bibitem{Olsh} M.A. Olshanetsky, "Supersymmetric two-dimensional
Toda lattice", {\it Commun. Math. Phys.} {\bf 88} (1983) 63.

\bibitem{LSS}  D.A. Leites, M.V. Saveliev, and V.V. Serganova,
"Embeddings of $osp(1|2)$ and the associated nonlinear
supersymmetric equations", in Group Theoretical Methods in
Physics, Vol. {\bf I} (Yurmala, 1985), VNU Sci. Press, Utrecht,
1986, 255.

\bibitem{And} V.A. Andreev, "Odd bases of Lie superalgebras and
integrable systems", {\it Theor. Math. Phys.} {\bf 72} (1987) 758.

 \bibitem{Ik} K. Ikeda, "A supersymmetric extension of the Toda
lattice hierarchy", {\it Lett. Math. Phys.} {\bf 14} (1987) 321.

\bibitem{EH} J. Evans and T. Hollowood, "Supersymmetric Toda field theories",
 {\it Nucl. Phys. } {\bf B 352} (1991) 723.


 \bibitem{LSor1}
O. Lechtenfeld and A.S. Sorin, "Fermionic flows and tau function
of the $N=(1|1)$ superconformal Toda lattice hierarchy", {\it
Nucl. Phys.}
  {\bf B 557} (1999) 535; "Hidden $N=(2|2)$ supersymmetry of the
$N=(1|1)$ supersymmetric Toda lattice hierarchy", {\it J. Nonlin.
Math. Phys.} {\bf 8} (2001) 183; "A note on fermionic flows of the
$N=(1|1)$ supersymmetric Toda lattice hierarchy", {\it J. Nonlin.
Math. Phys.} {\bf 11} (2004)
 294.


\bibitem{KS1}
V.G. Kadyshevsky and A.S. Sorin, "Supersymmetric Toda lattice
hierarchies", In "Integrable Hierarchies and Modern Physical
Theories" (Eds. H. Aratyn and A.S. Sorin), Kluwer Acad. Publ.,
Dordrecht/Boston/London, (2001) 289, nlin.SI/0011009.


\bibitem{CzJ} V.V.Gribanov, V.G. Kadyshevsky, and A.S. Sorin, "Periodic supersymmetric Toda lattice
hierarchy",
 {\it Czech. J. Phys.} {\bf 54}  (2004) 1289.

 \bibitem{DDNS} V.V.Gribanov, V.G. Kadyshevsky, and A.S. Sorin, "Generalized Fermionic
Discrete Toda Hierarchy", {\it Discrete Dynamics in Nature and
Society}  {\bf 2004:1} (2004) 113.

\bibitem{DLS}  V.B. Derjagin, A.N. Leznov, and A.S. Sorin,
  "The solution of the $N=(0|2)$ superconformal f-Toda lattice", {\it Nucl.Phys.}  {\bf B 527} (1998)
  643.


\bibitem{Carlet} G. Carlet, "The Hamiltonian Structures of the Two-Dimensional
Toda Lattice and R-matrices", math-ph/0403049.


\bibitem{KS2}
V.G. Kadyshevsky and A.S. Sorin, "N$=(1|1)$ supersymmetric
dispersionless Toda lattice hierarchy", {\it Theor. Math. Phys.}
{\bf 132} (2002) 1080; "Continuum limit of the $N=(1|1)$
supersymmetric Toda lattice hierarchy", JHEP Proceedings, PrHEP
unesp2002, Workshop on Integrable Theories, Solitons and Duality,
1-6 July 2002, Sao Paulo, Brazil.


\bibitem{KulSk}
P.P. Kulish and  E.K. Sklyanin, "On the  solution of the
Yang-Baxter equation", {\it  J. Sov. Math.} {\bf 19} (1980) 1596;
{\it Zap. Nauch. Sem. LOMI} {\bf 95} (1980) 129 (in Russian).



 \bibitem{GKS} V.V.Gribanov, V.G. Kadyshevsky, and A.S. Sorin,
 "Hamiltonian structures of fermionic two-dimensional Toda lattice hierarchies",
    nlin.SI/0505039.




\bibitem{AB} H. Aratyn and K. Bering, "Compatible Poisson Structures of Toda Type
 Discrete Hierarchy", nlin.SI/0402014.

\bibitem{STSh} M.A. Semenov-Tyan-Shanskii, "What a Classical
R-matrix Is", {\it Functsional. Anal. i Prilozhen.} {\bf 17}
(1983) 17 (in Russian); {\it Functional Anal. Appl.} {\bf 17}
(1983) 259.

\bibitem{Yung} C.M. Yung, "The modified classical Yang-Baxter equation and supersymmetric Gel'fand-Dikii brackets",
{\it Mod. Phys. Lett.} {\bf 8} (1993) 129.

\bibitem{OR} W. Oevel and O. Ragnisco, "R-matrices and Higher
Poisson Brackets for Integrable systems",  {\it Phys.} {\bf A 161}
(1989) 181.

\bibitem{LCP} L.-C. Li  and S. Parmentier, "Nonlinear Poisson Structures and
r-matrices", {\it Comm. Math. Phys.} {\bf 125} (1989) 545.

\bibitem{KaSa} K. Kajiwara and J. Satsuma, "The conserved
quantities and symmetries of the two-dimensional Toda lattice
hierarchy", {\it J. Math. Phys.} {\bf 32} (1991) 506.



\bibitem{Isaev}
 A.P. Isaev, "Quantum groups and Yang-Baxter equations",
{\it Sov. J. Part. Nucl.} {\bf 26} (1995) 501; {\it Fiz. Elem.
Chastits i At. Yadra} {\bf 26} (1995) 1204 (in Russian).


\end{thebibliography}
\end{document}